\documentclass[sigconf]{acmart}

\usepackage{lipsum}             
\usepackage[utf8]{inputenc}

\usepackage{amsmath,amsfonts}
\usepackage{graphicx}
\usepackage{array}
\usepackage{tabularx}
\usepackage{multirow}
\usepackage{pgfplots}
\usepackage{svg}
\usepackage[ruled]{algorithm2e}
\DeclareUnicodeCharacter{2212}{−}
\usepgfplotslibrary{groupplots,dateplot}
\usetikzlibrary{patterns,shapes.arrows}
\pgfplotsset{compat=newest}

\newtheorem{definition}{Definition}

\setcopyright{none}
\copyrightyear{2023}
\acmYear{2023}
\acmDOI{XXXXXXX.XXXXXXX}
\acmISBN{}
\acmConference[PoPETS]{Proceedings on Privacy Enhancing Technologies}{July 10--14, 2023}{Lausanne, Switzerland}
\begin{document}

\title{On the Privacy Risks of Deploying Recurrent Neural Networks in Machine Learning Models}

\author{Yunhao Yang*}
\email{yunhaoyang234@utexas.edu}
\affiliation{%
  \institution{University of Texas at Austin}
  \country{USA}
 }
 
 \author{Parham Gohari*}
\email{pgohari@utexas.edu}
\affiliation{%
  \institution{University of Texas at Austin}
  \country{USA}
 }
 
 \author{Ufuk Topcu}
\email{utopcu@utexas.edu}
\affiliation{%
  \institution{University of Texas at Austin}
  \country{USA}
}

\begin{abstract}
{
We study the privacy implications of training recurrent neural networks (RNNs) with sensitive training datasets.
Considering membership inference attacks (MIAs)—which aim to infer whether or not specific data records have been used in training a given machine learning model—we provide empirical evidence that a neural network's architecture impacts its vulnerability to MIAs.
In particular, we demonstrate that RNNs are subject to a higher attack accuracy than feed-forward neural network (FFNN) counterparts.
Additionally, we study the effectiveness of two prominent mitigation methods for preempting MIAs, namely weight regularization and differential privacy.
For the former, we empirically demonstrate that RNNs may only benefit from weight regularization marginally as opposed to FFNNs.
For the latter, we find that enforcing differential privacy through either of the following two methods leads to a less favorable privacy-utility trade-off in RNNs than alternative FFNNs: 
(i) adding Gaussian noise to the gradients calculated during training as a part of the so-called \textsc{DP-SGD} algorithm and (ii) adding Gaussian noise to the trainable parameters as a part of a post-training mechanism that we propose.
As a result, RNNs can also be less amenable to mitigation methods, bringing us to the conclusion that the privacy risks pertaining to the recurrent architecture are higher than the feed-forward counterparts.
}

\end{abstract}


\maketitle
\def\thefootnote{*}\footnotetext{These authors contributed equally to this work}

\section{Introduction} \label{sec: intro}

In the emerging applications of artificial intelligence, machine learning models are frequently trained with personal, proprietary, operational, confidential, or otherwise sensitive datasets, which raises privacy concerns.
Even when these datasets are securely stored and safeguarded from unauthorized access, sharing the outputs of a machine learning model that has been trained with such data can lead to unintended information leakage \cite{mireshghallah2020privacy,rigaki2020survey}.
Hence, it is imperative to foresee and preempt the privacy risks of training machine learning models with sensitive datasets.

We study the privacy risks of machine learning models that are powered by recurrent neural networks (RNNs) and compare them their counterparts powered by feed-forward neural networks (FFNNs). 
As opposed to FFNNs, in which the nodes in every layer are only connected to the nodes in the subsequent layers, RNNs allow for backward connections in their architecture.
RNNs are widely used in sequential machine learning tasks such as natural language processing \cite{wu2016google}, speech and handwriting recognition \cite{sak2014long,li2015constructing,graves2008novel}, deep reinforcement learning \cite{li2015recurrent, Liu_2017}, and semantic segmentation of video sequences \cite{Pfeuffer2019SemanticSO}.
While the privacy risks of neural networks—irrespective of their architecture—have been subject to an active line of research, an account of whether or not the architecture of neural networks affects their privacy risks remains unknown.

We consider \textit{membership inference attacks (MIAs)} as the underlying privacy threat. 
In an MIA, an adversary is allowed to query the output of a neural network with a collection of data records, and must subsequently infer whether or not those data records belong to the neural network's training dataset \cite{hu2021membership}.
Successful instances of these attacks with minimal access to the neural network can have significant privacy ramifications for the individuals who populate the training datasets with their data. For example, consider a machine learning model that has been trained with the data of individuals with certain characteristics such as a particular ethnic origin, religion, medical condition, gender, or sexuality. In this case, a successful MIA that asserts—or refutes—the membership of an individual's data can reveal such sensitive characteristics.

Our contributions in this paper are twofold: 
the first contribution concerns how RNNs and FFNNs compare in vulnerability to MIAs and the second contribution concerns defending them against MIAs.

In the first contribution, we design and conduct a series of experiments to compare RNNs and FFNNs in their vulnerability to MIAs in three representative machine learning tasks, namely image classification, machine translation, and deep reinforcement learning. In order to study the impact of network architecture on vulnerability to MIAs, we are mindful to separate other known influential factors such as overfitting \cite{salem2018ml,Shokri_2017,Yeom2018PrivacyRI,bentley2020quantifying}, number of trainable parameters \cite{nasr2019comprehensive}, diversity of the training data \cite{long2018understanding}, and number of prediction classes \cite{truex2019demystifying}.
Taking all of these factors into account, we observe that the MIAs consistently achieve a higher attack accuracy against the RNN models.

In order to investigate the root causes of the observed higher vulnerability in RNNs, we further study the behavior of the two models when they are queried with members of their training datasets and unseen data.
We observe that when the uncertainty of the two models' predictions in terms of entropy is equal with respect to the validation dataset, the entropy with respect to the training data is lower in RNNs.
Moreover, we demonstrate that the decisions of the MIAs resemble establishing a threshold for prediction entropy to distinguish member data from non-members.
In such a threshold-based inference, a larger gap between the entropy of the predictions in training and validation—as observed in RNNs—increases attack accuracy.
Moreover, we demonstrate that subsequent gradient updates in FFNNs can mask the membership of data used early in the history of training, whereas the MIAs remain relatively accurate even for such outdated data in RNNs.

In the second contribution, we shift the focus from vulnerability analysis to mitigation methods. A popular mitigation approach is to prevent overfitting as a root cause of vulnerability to MIAs—most prominently via weight regularization \cite{Shokri_2017, salem2018ml}.
While weight regularization have been shown to be oftentimes highly effective for FFNNs to preempt MIAs, in the experiments we demonstrate that the RNN models benefit from regularization only marginally as opposed to the FFNN models. As a result, RNNs may be not only more vulnerable to MIAs, but also harder to be defended against them.

Methods that leverage the promise of differential privacy are known to be the most effective in defending neural networks against MIAs \cite{hu2021membership}. However, the protection afforded by these methods typically comes at the expense of a reduction in utility in terms of model performance \cite{rahman2018membership}. These methods impose an error margin on the inference power of MIAs and the error can be balanced against utility loss through adjusting the level of differential privacy \cite{Yeom2018PrivacyRI}.

Existing methods enforce differential privacy by obfuscating either of the following:
the objective function \cite{zhang2012functional} during training, the gradients calculated during training \cite{abadi2016deep,brendan2018learning}, or the model's parameters post training \cite{chaudhuri2011differentially,wu2017bolt,lu2022differentially}. 
Post-training methods offer more flexibility in adjusting the level of differential privacy because, in order to adjust the level of privacy in the first two methods, the model must be retrained from scratch.
On the other hand, post-training methods may be less advantageous with respect to the privacy-utility trade-off that they face \cite{abadi2016deep}.

We compare RNNs and FFNNs in their utility loss due to differential privacy considering two representative enforcement methods: the celebrated \textsc{DP-SGD} algorithm \cite{abadi2016deep} which adds Gaussian noise to the gradients during training and a mechanism that we develop which adds Gaussian noise to the trained parameters of a neural network post training.
For both methods, the experiment results indicate that adding the same level of noise degrades more utility in RNNs than FFNNs.

The proposed post-training mechanism may be of independent interest.
We show that the mechanism satisfies a relaxation of differential privacy called \textit{random differential privacy} \cite{Hall_Wasserman_Rinaldo_2013}.
In noise-additive differential privacy mechanisms, the noise is typically calibrated with the extent to which a single record of the training dataset can change the model's outcome, formally called \textit{sensitivity}. 
Computing sensitivity analytically can be very challenging and one might have to resort to an upper bound for the sensitivity which can be too loose and subsequently cause too much noise to be added \cite{chaudhuri2011differentially,wu2017bolt,lu2022differentially}.
Alternatively, random differential privacy fixes a level of sensitivity with some confidence level and guarantees differential privacy for data records that give rise to that level of sensitivity. 

We use the \textsc{SensitivitySampler} algorithm \cite{rubinstein2017pain} to estimate the sensitivity of the models.
We show that by utilizing these estimates, we are able to achieve an acceptable privacy-utility trade-off for the models in the experiments: reducing the MIAs' attack accuracy to roughly 50\%—equivalent to random guessing—while trading off less than 10\% utility.
We further observe that the sensitivity estimates for the RNN and FFNN models in the experiments take similar values.
As a result, adding the same level of noise to the two models using the \textsc{DP-SGD} algorithm and the proposed post-training mechanism, satisfies the same level of conventional and random differential privacy, respectively, yet it leads to more utility loss in RNNs than FFNNs.
Since RNNs were consistently rendered more vulnerable to MIAs and more difficult to be defended, this paper provides strong empirical evidence that the privacy risks of RNNs are more severe than FFNNs.

\section{Preliminaries} \label{sec: prelim}

In this section, we first review some background about the differences between RNNs and FFNNs. Then, we introduce the machine learning tasks that we consider in the experiments.

\subsection{Recurrent vs. Feed-Forward Architecture}
A neural network comprises a collection of nodes, each of which accepts an input and produces an output according to a fixed mapping called an activation function. 
The architecture of a neural network determines how the nodes of the networks are connected to one another.
In a feed-forward architecture, the nodes can be stacked into an ordered sequence of layers from the network's input to its output such that the output of each node only affects the nodes in the subsequent layers.
Examples of FFNNs include multi-layer perceptrons (MLPs), convolutional neural networks (CNNs), and more sophisticated designs such as transformers \cite{vaswani2017attention}.

In a recurrent architecture, the connections between the nodes may form a cycle.
The backward connection between an RNN's nodes can be unfolded into an infinite sequence of layers, each of which represents the node activations at different time steps. 
Therefore, an RNN's output depends on the entire history of its inputs, which results in exhibiting a temporally dynamic behavior. Such features make RNNs suitable for processing sequential data such as sentences, videos, and audios \cite{dupond2019thorough}.

FFNNs can also exhibit a temporally dynamic behavior through cascading MLPs or CNNs, or, more intelligently in transformers; however, the outputs in these methods only depend on a finite window in the history of inputs. Therefore, RNNs appear to be more expressive than FFNNs.
On the other hand, it easier to parallelize the training of FFNNs \cite{gehring2017convolutional}. For example, generative pre-trained transformers \cite{radford2018improving} leverage parallelization to train natural-language processing models on very large datasets.
Furthermore, making predictions based on the entire history of inputs may be unnecessary as theoretically shown in \cite{10.1145/3188745.3188954}.
As a result, there has been an increasing interest—with many successful instances—in replacing RNNs with FFNNs \cite{dauphin2017language, vaswani2017attention, gehring2017convolutional, miller2018stable}.

\subsection{RNN Applications Considered}
In the experiments, we consider three representative machine learning tasks: image classification, machine translation, and deep reinforcement learning. In the sequel, we briefly introduce each of the tasks. Then, we state some of the possible privacy harms that MIAs may cause specific to these tasks.

In the image classification task, the model must label a given image using a fixed set of classes. 
The model's output is a probability vector that determines its confidence in assigning each of the labels to the given image.
CNNs are dominantly used in image classification \cite{Krizhevsky2012ImageNetCW, Gavrikov2022CNNFD, Graham2014SpatiallysparseCN}; however, RNNs may also be used to process images as a sequence of pixels \cite{visin2015renet}.

In image classification tasks, the models may be trained with labeled image dataset that contain sensitive information.
For example, consider a healthcare provider who fine-tunes a medical image classifier to predict the risk factors pertaining to a certain disease for a certain minority population. In this case, an MIA with access to an aggregated list of patient records can infer whether or not some of the data subjects belong to the considered minority group.



In the machine translation task, the model must map a sequence of words, syllables, or otherwise tokens from a fixed input dictionary to a target dictionary.
Both RNN and FFNN solutions use an \textit{encoder-decoder} framework. 
The first half of the model—the encoder—computes an encoding of the input sequence through multiple encoder layers.
Subsequently, the second half of the model—the decoder—uses the encoding and generates an output sequence through multiple decoder layers.
The output of the model is a sequence of probability vectors over the target dictionary words. The dictionaries are appended with start and end tokens to signal the start and completion of sentences, respectively.

Recurrent architectures such as bi-LSTM \cite{wu2016googles} use a network of long-short term memory (LSTM) units to construct both of the encoder and the decoder networks and can process arbitrary-length sequences.
Feed-forward architectures such as transformers \cite{vaswani2017attention} fix a window of sequence lengths allowed and construct the encoder and the decoder networks using FFNNs.
Both of the above network architectures are widely used in machine translation; however, since the debut of transformers in 2017, they have outperformed RNN-based models \cite{wolf-etal-2020-transformers}.

As for the privacy risks that MIAs can pose to machine translation models, assume that some business analytics tool trains a model using internal meeting transcripts as training datasets. In this case, an MIA can infer whether or not a given sentence has been discussed in the meetings.  

Finally, in the deep reinforcement learning task that we consider, an agent must learn how to navigate through an unknown map and reach a target state under partial state observations.
At every state observation, the agent must compute a probability vector over the available actions, which is called a policy.
Under partial state observations, the optimal policy may require memory \cite{lusena2001finite} and RNNs can be integrated with deep reinforcement learning algorithms to capture long-term dependencies \cite{chen2016deep}.
Both MLPs and LSTMs are commonly used in deep reinforcement learning algorithms such as soft actor critic \cite{haarnoja2018soft}, proximal policy optimization (PPO) \cite{schulman2017proximal}, etc.

In deep reinforcement learning tasks, an MIA can infer whether or not specific locations have been used to train the agent. For example, a new owner of an autonomous vehicle may be able to infer whether or not the previous owner has visited certain locations, thereby violating the previous owner's location privacy.

\section{Methods} \label{sec: method}
In this section, we describe the threat model that we consider for MIAs. Then, we lay out the methodology that we use to design the MIAs in the experiments and compare the MIA layouts with existing MIAs in the literature.

\subsection{Threat Model and Assumptions}
There exist two parties in the threat model that we consider: a victim and an attacker.
The victim aims to train a neural network for a given machine learning task and a given training dataset. For example, the victim may train an image classifier using a dataset of labeled images.
In order to train the neural network, the victim must choose a training algorithm alongside its hyperparameters, loss function, and the neural network specifications—including the number of hidden layers, number of nodes per layer, architecture, activation functions, etc.
Once the victim's neural network is fully trained, the victim proceeds with generating predictions for outsider inquiries, i.e., the victim receives a data record and subsequently responds by publishing its predictions for the received data.

\begin{figure}
    \centering
    \includegraphics[width=\columnwidth]{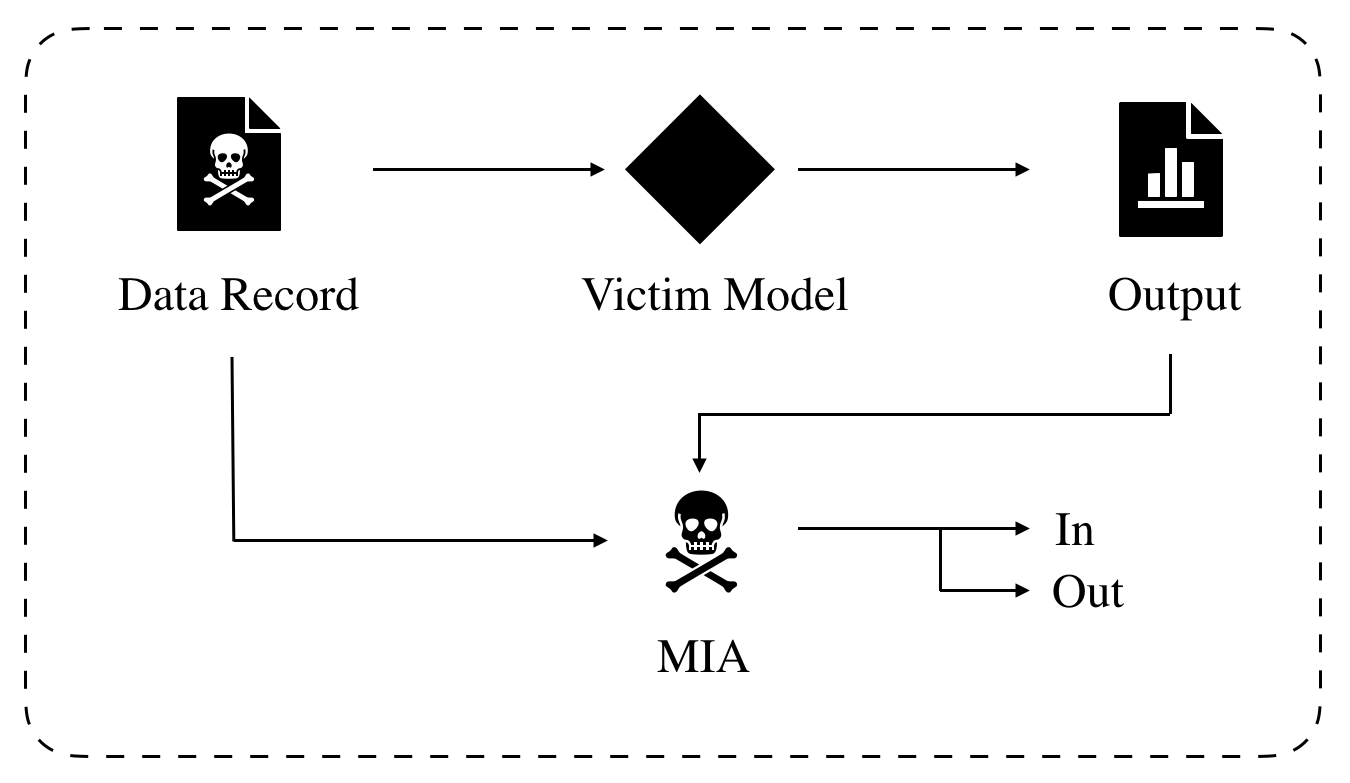}
    \caption{Schematic of the execution of an MIA.}
    \label{fig:MIA}
\end{figure}

The attacker in the considered threat model conducts an MIA; that is, it submits an inquiry to the victim using some data record and must infer whether or not the data record belongs to the victim's training dataset as depicted in Figure \ref{fig:MIA}.
MIAs are typically categorized into two groups: black-box and white-box attacks \cite{hu2021membership}. The former group assumes that the attacker can only access the input and the output of the victim's neural network. White-box attackers may have access to the value of the weights and the output of the nodes anywhere in the victim's neural network. Additionally, white-box attacks may also probe the victim's loss function and its gradient for the queried data record.

In terms of the side-information that is available to the attacker, the survey in \cite{hu2021membership} assumes that a black-box MIA's side-information is limited to the distribution of the training data—implying that the attacker can obtain a training dataset that is similar to that of the victim.
The survey considers any additional assumption on the available side-information as an indicator of white-box attacking.
However, such a distinction about side-information is not uniformly followed in the literature; for example, the work in \cite{Shokri_2017} assumes that the attacker knows the training algorithm of the victim, and the MIAs in \cite{Sablayrolles2019WhiteboxVB} are provided with side-information about the victim's training algorithm, hyperparameters, and the network specifications, yet both works consider their MIAs as black-box attacks.
Access to the value of the loss function and its gradient is the key enabler that enhances the attack accuracy in white-box attacks as empirically demonstrated by \cite{nasr2019comprehensive}. 
We therefore draw the line between black-box and white-box attacks based on the access granted to the attacker and not the side-information.

We now state our assumptions on the attacker's access limits and side-information.
In terms of access limitations, we assume that the attacker has a black-box access to the input and output layers of the victim's neural network. 
The attacker is able to access the input layer via submitting an unlimited number of inquiries and is able to observe the output layer via evaluating the confidence scores with which the victim responds to an inquiry.
In terms of the side-information that is available to the attacker, we assume that the attacker knows the victim's task, training algorithm alongside its hyperparameters, and the network specifications.

The authors of \cite{Sablayrolles2019WhiteboxVB} show that an optimal MIA—under mild assumptions on the distribution of the neural network's parameters—utilizes the victim's full confidence scores. 
Our goal in the experiments is to investigate whether or not the architecture of a neural network affects its vulnerability to MIAs.
As a result, we assume full confidence-score observability in the threat model in an effort to maximize the accuracy of the MIAs and control the variables that have the potential of affecting the accuracy of MIAs besides architecture.

\subsection{Designing MIAs}
\begin{figure}[t]
    \centering
    \includegraphics[width=\columnwidth]{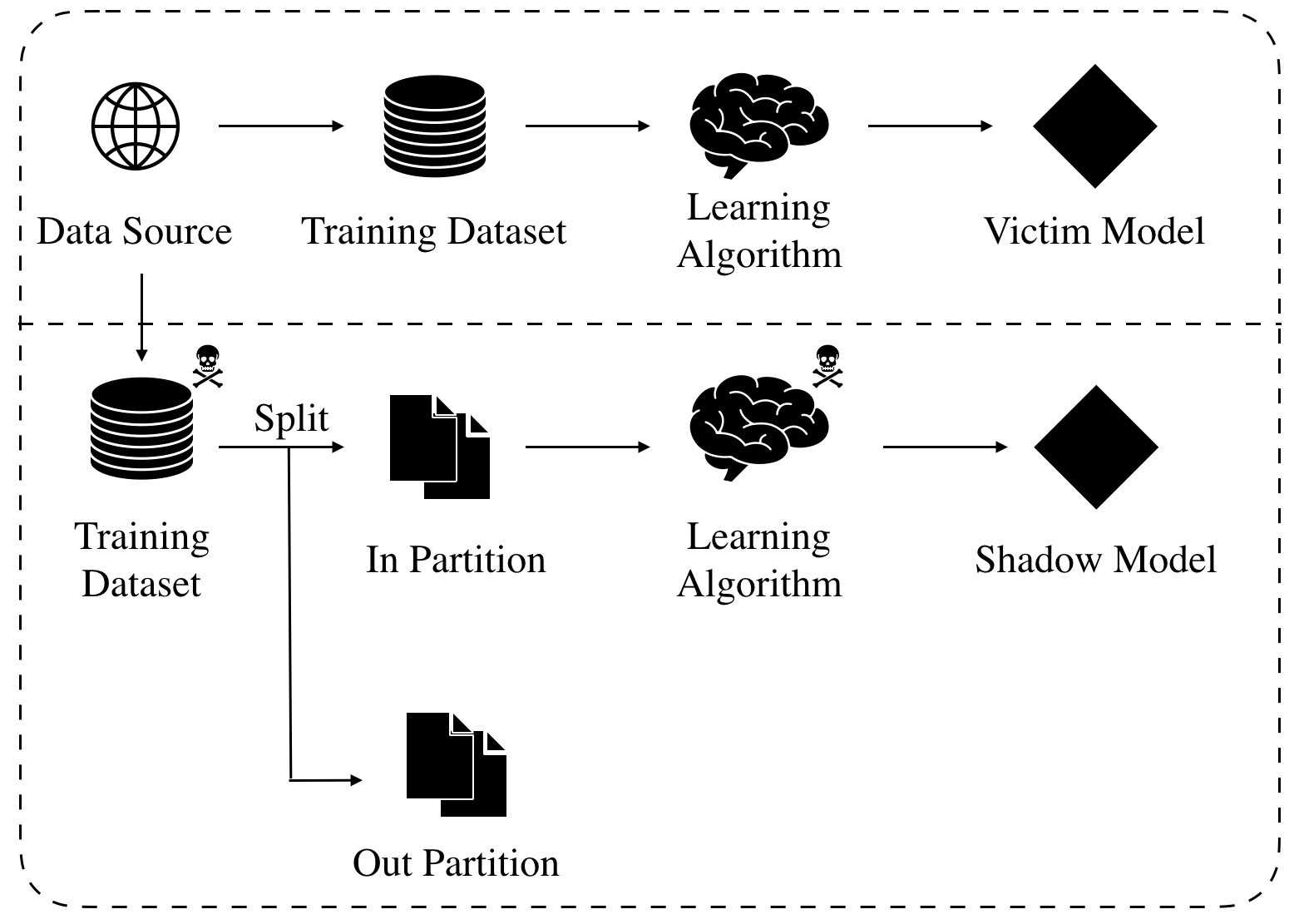}
    \includegraphics[width=\columnwidth]{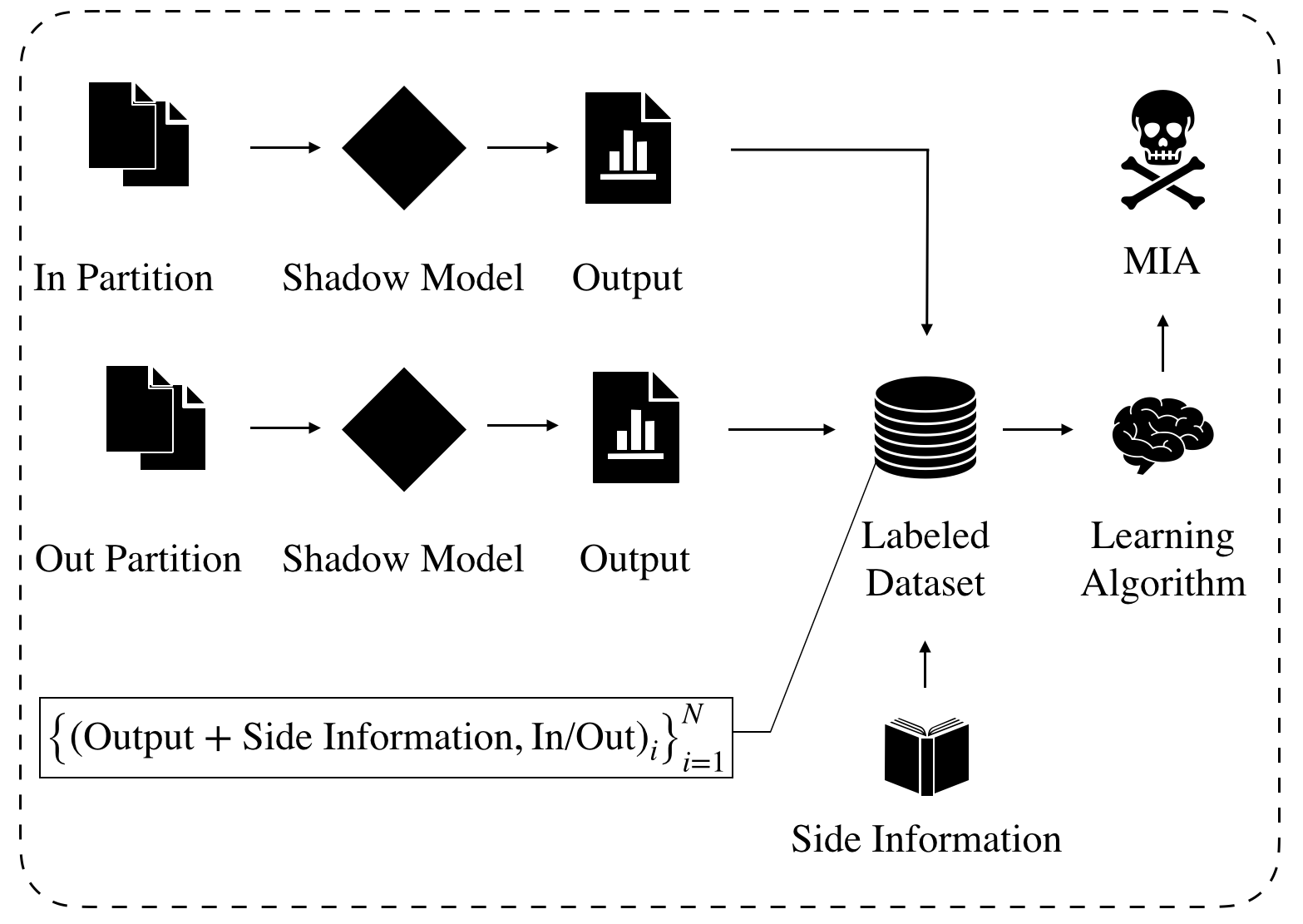}
    \caption{Top: a schematic of training shadow models. The skull above the shadow model's training dataset and learning algorithm indicates that the two components may differ between the victim and the attacker's side. Bottom: training the MIA using the output of the shadow model and side-information.}
    \label{fig:training MIA}
\end{figure}

We follow the framework of \textit{shadow models} \cite{Shokri_2017} in designing the MIAs in this paper.
Intuitively, a shadow model must mimic the victim's behavior without having access to its training dataset.
Following the assumption that the attacker knows the victim's training data distribution, we assume that there exists a data source from which the attacker can obtain a similar training dataset to train a shadow model—see Figure \ref{fig:training MIA}.
In order to increase accuracy, MIAs often obtain multiple training datasets from the data source and subsequently train multiple shadow models to better mimic the victim's behavior.

For the image classification and machine translation tasks, we allocate two disjoint partitions of a large dataset to the victim and the attacker, separately.
Analogously for the deep reinforcement learning task, we use two disjoint sets of environment maps for the victim and the attacker to train their models.

In the next step, the attacker splits its training dataset into two partitions.
The first partition will be used to train the shadow models following the side-information that the attacker has regarding the victim's training algorithm and network specifications.
Once the shadow models are trained, the attacker is provided with a proxy to the victim's neural network.
The attacker knows which data records it has used to train the shadow models and it knows that the second partition has not been used to train any of the shadow models.
Therefore, the attacker can train a binary classifier to distinguish between the outputs that correspond to member data and those corresponding to non-member data.
The attacker can then use the binary classifier against the victim to execute the MIA as depicted in Figure \ref{fig:MIA}. 

Shadow-model-based MIAs typically use a 3-tuple format for the entries of the binary classifier's training dataset as shown in the bottom box in Figure \ref{fig:training MIA}.
Each entry corresponds to a query that is made from a shadow model and the query may originate from either partitions of the attacker's training dataset.
The first element contains the shadow model's output, the second element indicates what the shadow model's optimal output must have been, and the third element indicates whether the query was made using a member data record or a non-member data record.

For the first element, we use the shadow model's full confidence score in vector format.
If the shadow model's output is a sequence of predictions—as the case in machine translation and deep reinforcement learning—we concatenate the confidence scores into one vector.

For the second element, we use the query's corresponding label in the training dataset as a one-hot vector or a concatenation of a sequence of one-hot vectors—such labels are readily available in the image classification and machine translation tasks.
In the deep reinforcement learning task, such a labeled training dataset does not exist; we therefore train a ``labeling agent'' to generate these labels.
The labeling agent simply learns a reward-maximizing policy in the environment in which a shadow model's policy is queried.
For each state observation at which the shadow model's policy is queried, the labeling agent provides its policy as a reference label.

Once the binary classifier's training dataset is fully populated, the attacker uses the first two elements as features and use the third element as binary labels—member or non-member—and concludes the design of the MIA by training a binary classifier that distinguishes between member and non-member queries.
The attacker then uses the trained binary classifier against the victim to execute the MIA as shown in Figure \ref{fig:MIA}.

\subsection{Connection with the Existing MIAs}
We now state how the MIAs implemented in this paper compare with the existing MIAs in the literature.
Regarding the image classification task, our MIA design is identical to the design in \cite{Shokri_2017}.
However, for the machine translation and deep reinforcement learning experiments, the existing MIAs have minor incompatibilities with this paper's threat model which we address by modifying them.

The works in \cite{song2019auditing,10.1162/tacl_a_00299} study developing MIAs specifically for machine translation models.
However, neither of the two existing works assume full confidence-score observability because they both aim to design practical MIAs with minimal side-information assumptions. 
In particular, the authors of \cite{song2019auditing} develop an MIA that is intended to be used by individuals who wish to \textit{audit} a natural language processing model—a process by which the individuals investigate whether or not their data has been used to train a natural language processing model.
In this scenario, full confidence-score observability is not realistic and the authors feed a redacted list of the output word rankings to the MIAs, instead.
We use the same MIA design as \cite{song2019auditing} except for that we use full confidence scores instead of word rankings.

The authors of \cite{10.1162/tacl_a_00299} use a similar design, but they take a further step towards developing practical MIAs and drop the assumption that the MIA knows the underlying distribution of the training data. However, the authors find that the resulting MIAs are not effective as their accuracy does not exceed random guessing by much.




We now review the existing MIAs in deep reinforcement learning tasks. The work in \cite{pan2019you}—which we follow closely in our MIA design for deep reinforcement learning—is the first to consider a privacy attack against reinforcement learning agents that resembles MIAs. However, instead of modeling the MIA as a binary classifier, the privacy attack uses a multi-class classifier. As a result, the attacker must train a labeling agent for every possible environment map prior to the execution the attack. By using a binary classifier, we train labeling agents only for the environment map with which the MIA is faced.
In another work, the authors of \cite{gomrokchi2021did} develop an MIA that infers the membership of a batch-constrained deep Q-learning agent's roll-out trajectories stored in its replay buffer. We do not follow the above work's methodology because we do not restrict the algorithm that is used to train the reinforcement learning agents.

\section{Vulnerability to Privacy Threats} \label{sec: vuln}


In this section, we report and analyze the results of a series of experiments by which we compare the vulnerability of RNNs and FFNNs to MIAs. 
In order to perform a meaningful comparison, we must control factors that affect vulnerability to MIAs other than network architecture.
We review these vulnerability factors and discuss how we take them into account in our experimental setup. 
Finally, we report and analyze the numerical results.

\subsection{Vulnerability Factors}
Overfitting has been extensively studied as the main source of vulnerability of machine learning models to MIAs \cite{hu2021membership}. 
Overfitting refers to the condition in which a machine learning model performs poorly when queried with data records outside its training dataset.
There exist mounting empirical evidence that MIAs are more successful against models that overfit their training data \cite{salem2018ml,Shokri_2017}.
However, there also exist successful instances of MIAs used against models with relatively low overfitting \cite{long2018understanding}.
In these instances, the underlying distribution of the training data as well the size of the training datasets may leave some data records more vulnerable than others. By identifying such data records, an MIAs may still maintain a high attack accuracy for models with low overfitting  \cite{hu2021membership}.

A theoretical account of the connection between overfitting and MIA accuracy remained unknown until the work in \cite{Yeom2018PrivacyRI}. The said work characterizes overfitting by \textit{average generalization error} defined as
\begin{equation}\label{eq:generalizaiton power}
    R_\text{gen} = \mathop{\mathrm{E}}\limits_{\substack{S\sim \mathcal{D}^n \\ z\sim\mathcal{D}}}[\ell_S(z)] - \mathop{\mathrm{E}}\limits_{\substack{S\sim \mathcal{D}^n \\ z\sim\mathcal{S}}}[\ell_S(z)],
\end{equation}
where $\mathcal{D}$ is the underlying distribution of the training data; $S$ is the victim's training dataset comprising $n$ samples drawn from $\mathcal{D}$; $\ell$ is a fixed loss function; and $\ell_{S}(\cdot)$ is the value of the model's loss function after being trained with $S$.

Under the assumption that the victim's loss function is bounded above and its value is accessible to the attacker, Yeom et al. establish that a higher average generalization error is a sufficient condition—but not necessary—for a higher attack accuracy \cite{Yeom2018PrivacyRI}. The authors further provide empirical evidence that the sufficiency relationship holds when the assumptions are relaxed to black-box MIAs. Later, a theoretical account of the relationship between generalization gap—training accuracy minus validation accuracy—and the accuracy of black-box MIAs was provided in \cite{bentley2020quantifying}.

In light of the established relationship between overfitting and attack accuracy, we are mindful to consider RNNs and FFNNs with similar training and validation performance levels. In order to control the effect of the size and the distribution of the victim's training dataset on vulnerability to MIAs, we use the same training dataset for both of the RNN and FFNN models. As the vulnerability to MIAs may not be evenly distributed across a collection of data records \cite{long2018understanding}, we evaluate the MIAs against the RNN and FFNN models using the same dataset. Finally, we consider RNNs and FFNNs whose number of parameters are close because it has been empirically demonstrated that a higher number of parameters increases vulnerability to MIAs \cite{nasr2019comprehensive}. 

\subsection{Experimental Setup}
We consider three representative machine learning tasks for the experiments of this section: image classification, machine translation, and deep reinforcement learning. 
In image classification, consistent with the threat model, we assume that there exist a data source that generates labeled image samples and use the \texttt{CIFAR10} dataset \cite{krizhevsky2009learning} as $50,000$ samples drawn from the data source.
We split these samples evenly into two partitions: one used by the victim and the other used by the attacker for the training of the shadow models—we train 5 shadow models. 

With the victim's portion of the training samples, we separately train an FFNN model and an RNN model.
The FFNN model is an instance of ResNet101 \cite{He2016DeepRL} implemented in the Keras library \cite{chollet2015keras} and specified as follows: 101 convolutional layers followed by one max-pooling layer, one fully connected linear layer, and an output layer with softmax activation. For the RNN model, we use ReNet \cite{visin2015renet} implemented by PyTorch \cite{paszke2019pytorch} under default parameters, which has the following specifications: 4 bi-directional LSTMs, 2 fully connected layers with ReLU activation, and an output layer with softmax activation. The former model contains $42,678,666$ trainable parameters and the latter has $42,569,590$ trainable parameters. Both models use the categorical cross-entropy loss function as their learning's objective function and use the Adam optimizer. The learning rates used are $0.001$ and $0.01$ for the former and the latter model, respectively. 
The shadow model of the MIAs use the same specifications as the victims for their training.

For the machine translation experiments, we choose translation from French to English. Similar to image classification, we assume there exist a data source from which translated pairs of English and French sentences can be sampled. We take the \texttt{Multi30K} dataset \cite{multi30k} as $30,000$ samples from the data source and split the samples evenly between the victim and the attacker.

For the RNN model, we use a bi-directional LSTM with dot-product attention mechanism developed in \cite{luong2015effective}. For the FFNN model, we use the standard transformer network specified in \cite{vaswani2017attention}. The RNN model and the FFNN model have $3,213,191$ and $3,225,714$ trainable parameters, respectively. Both networks use the negative log likelihood function as their learning algorithm's loss function and use Adam optimizer with learning rate $0.001$.  

\begin{figure}[t]
    \centering
    \includegraphics[width = \columnwidth]{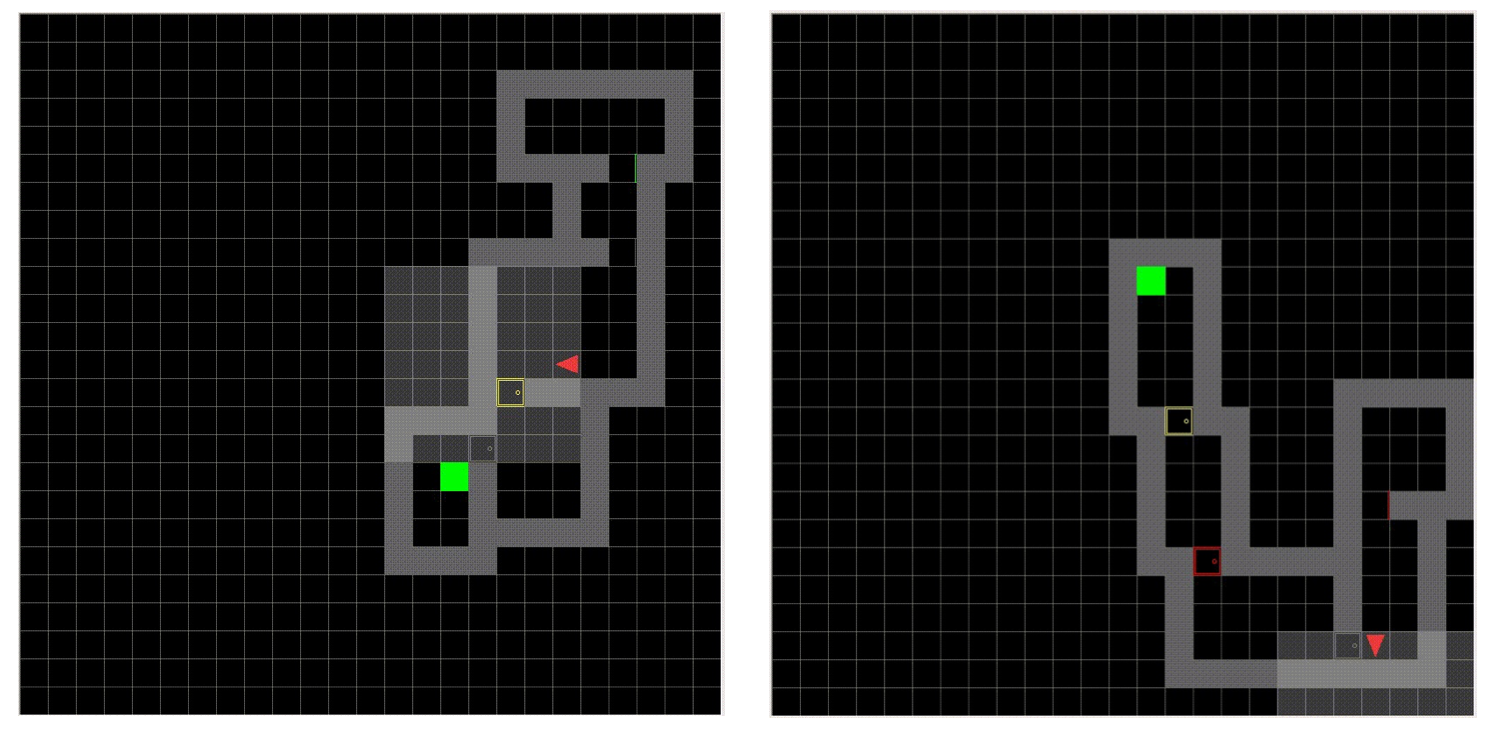}
    \caption{The MiniGrid-MultiRoom-N4-v0 environment. The agent (red triangle) must find the goal state (green square) while observing the highlighted box that surrounds it. The environment consists of multiple floor-maps, two of which are shown above.}
    \label{fig: DRL task}
\end{figure}

Finally, for the deep reinforcement learning task, we use the MiniGrid-MultiRoom-N4-v0 environment from the MiniGrid toolkit \cite{gym_minigrid}.
The victim's goal is to train a deep reinforcement learning agent that can navigate its way through four rooms with closed doors and reach the green tile as shown in Figure \ref{fig: DRL task}.
The victim is provided with a limited number of floor-maps for training and must generalize to unseen floor-maps. 
The attacker's goal, on the other hand, is to infer the membership of floor-maps.
In this experiment, the MiniGrid toolkit serves as the data source and the attacker is able to obtain an arbitrary number of floor-maps by feeding a randomly generated seed number to the toolkit's simulator. 

For the FFNN agent, we use an MLP network with $5,335$ trainable parameters and the following specifications: the actor network has two hidden layers with dimension 74, and the critic network has 2 hidden layers with dimension 64. Both networks use a softmax output layer and $\tanh$ as their activation function. The RNN agent uses an MLP with some additional LSTM cells. The RNN has $5,216$ trainable parameters and its specifications are as follows: both the actor and the critic networks have 2 linear layers with hidden dimension 32, 4 single-directional LSTM cells, and a softmax output layer. We use the PPO algorithm \cite{schulman2017proximal} implemented by the RL-Starter-Files library \cite{rl_starter} with default parameters to train the victim agents and their respective shadow models and labeling models.

\begin{figure*}[!ht]
    \centering
    \input{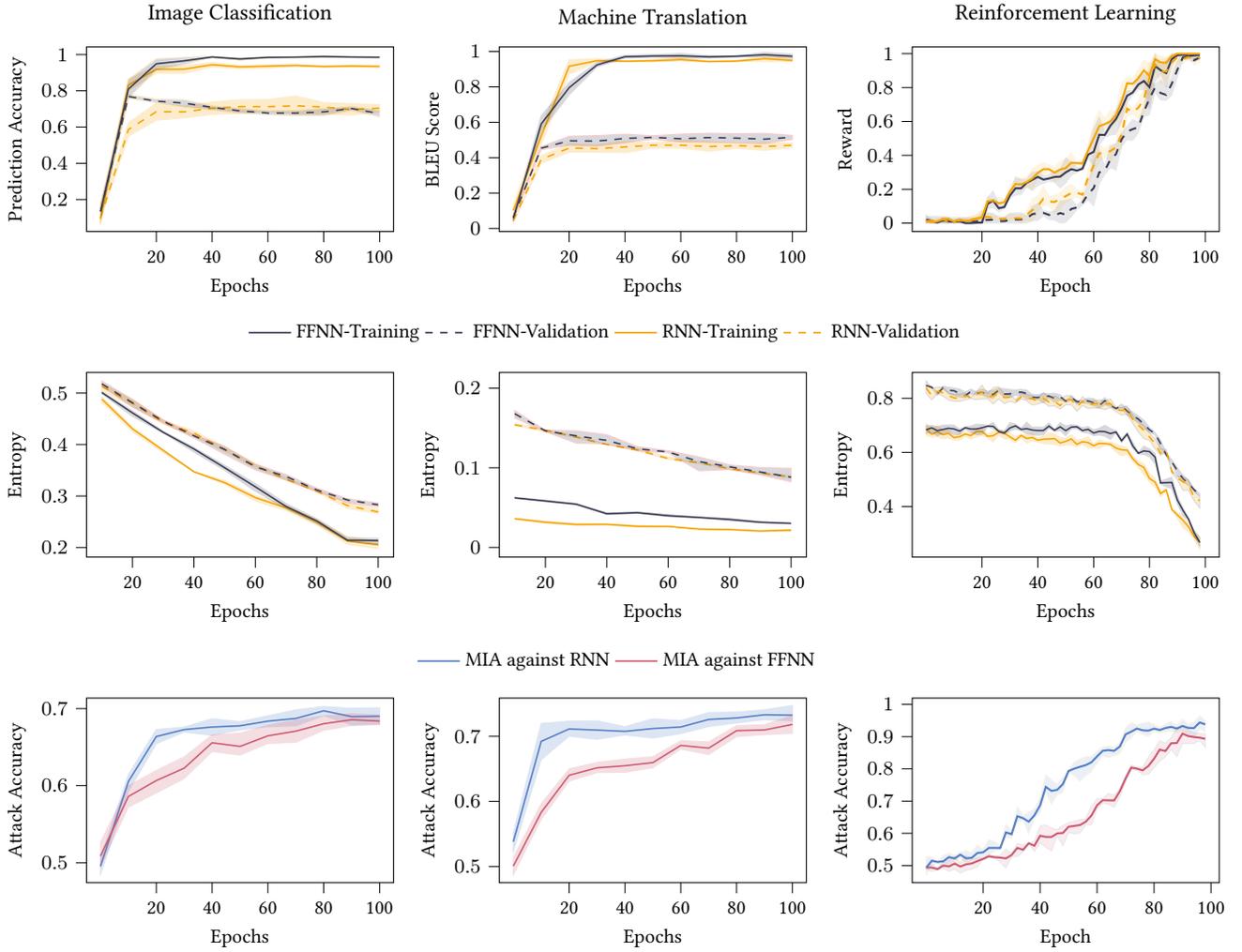}
    \caption{Comparing the vulnerability of RNNs and FFNNs to MIAs in three representative machine learning tasks: image classification (left), machine translation (center), and deep reinforcement learning (right). The first row plots training and validation performance vs. number of epochs; the second row plots the average prediction entropy vs. number of epochs; and the third row plots attack accuracy vs. number of epochs.}
    \label{fig:MIA accuracy}
\end{figure*}

\subsection{Numerical Results}
Following the experimental setup above, we train each of the described RNN and FFNN models for a range of epoch numbers and plot the training and the validation performance of the models. For the image classification experiment, we use the percentage of the correct predictions—or prediction accuracy—as the performance measure; for machine translation, we measure performance using the bilingual evaluation understudy (BLEU) score \cite{papineni-etal-2002-bleu}, which captures how a model's translation correlates to that of a human; and for deep reinforcement learning, we use the total episodic reward as the performance measure. The reward at time-step $t$ is
\begin{equation}
    r(t) = \begin{cases}1 - \gamma \frac{t}{T}, &\text{goal reached,}\\ 0, &\text{otherwise,}\end{cases}
\end{equation}
where $T$ is the episode length—set to $200$ in the experiments— and $\gamma$ is the discount factor—set to $0.9$.

At every epoch number tested, we train a separate MIA whose shadow models are trained for the same number of epochs as that of the victim.
We evaluate the performance of the MIAs by measuring the percentage of correct inferences which we refer to as attack accuracy.
In all instances, the MIAs' validation datasets have an equal number of members and non-member records; hence, random guessing in this case achieves $50\%$ attack accuracy.

In Figure \ref{fig:MIA accuracy}, it can be observed that the attack accuracy against the RNN models is consistently higher than it is against FFNN models. In particular, The RNNs are more vulnerable before and after the performance level of the victims converges; however, the gap between the attack accuracy of the two MIAs narrows as the models train for higher epoch numbers.

In the image classification experiment, the validation performance of the RNN and FFNN models are approximately equal. The FFNN model has a higher generalization gap than the RNN model upon convergence, and it has slightly more trainable parameters, yet surprisingly, the attack accuracy against the FFNN model is lower than the RNN model. In the machine translation experiment, the two models have approximately equal performance levels both in training and validation but the MIA against the RNN achieves a higher attack accuracy. Finally, in the deep reinforcement learning experiment, the two models appear to have a zero generalization gap upon convergence, yet the MIA against the RNN model is more accurate than it is against the FFNN model. 

\paragraph{\textbf{Prediction Entropy as a Vulnerability Factor:}}
In order to further investigate the reasons behind the excessive vulnerability of RNNs to MIAs, we measure the uncertainty of the models' outputs in terms of \textit{average prediction entropy}, which we define as follows: let $m$ be the number of prediction categories and $Y = \left\{\left(y^{(i)},p^{(i)}\right)\right\}_{i=1}^L$ be a sequence of $L$ pairs of prediction outcomes and confidence scores, respectively. Then, the corresponding average prediction entropy is
\begin{equation} \label{eq: RL reward function}
    H_\mathrm{APE}(Y) = -\frac{1}{L}\sum_{i=1}^L \sum_{j=1}^m p^{(i)}_j \log\left(p^{(i)}_j\right).
\end{equation}

\begin{figure*}[!ht]
    \centering
    \input{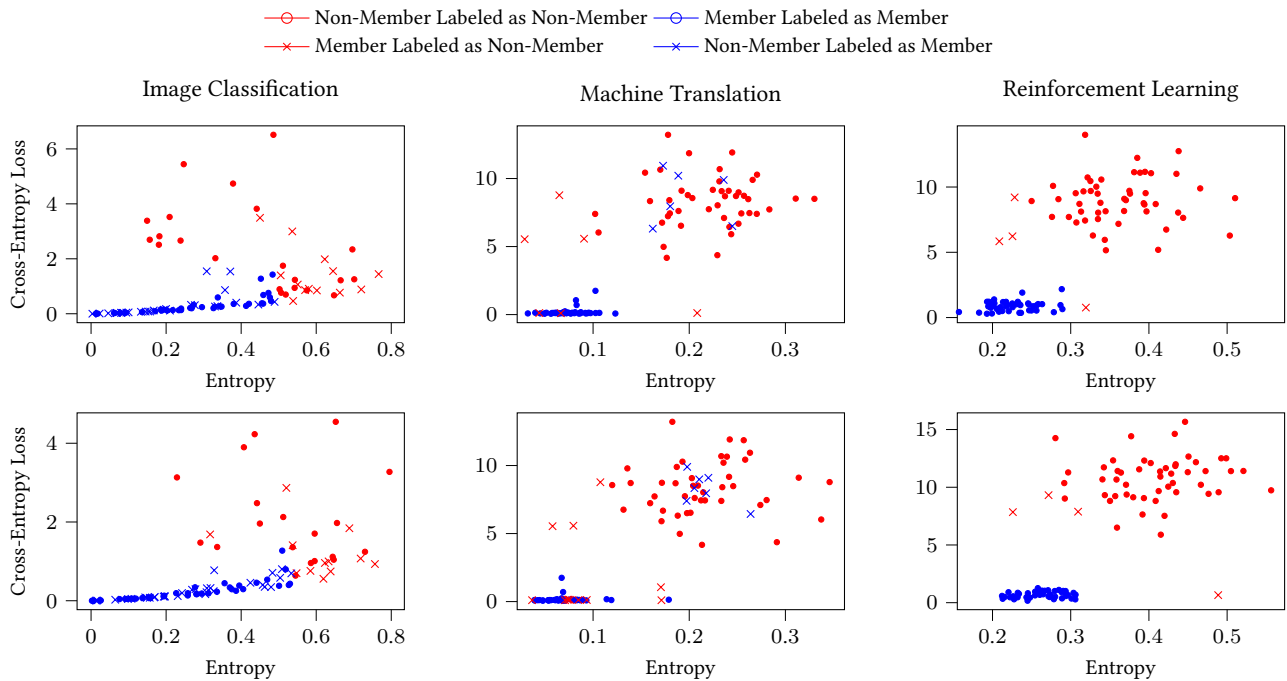}
    \caption{An illustration of the decision boundaries of MIAs with respect to cross-entropy loss and entropy of the confidence scores. The top row shows the results for RNNs and bottom row shows the results for FFNNs.}
    \label{fig:entropy_scatter}
\end{figure*}

We report the average prediction entropy of the RNN and the FFNN models in Figure \ref{fig:MIA accuracy}. The results show that, while the prediction entropy of the two models are approximately equal over the validation dataset, their prediction entropy with respect to their training data differ noticeably—at least in the early stages of training. In the initial stages, the entropy gap between the validation and the training dataset in RNNs is larger than FFNNs. As the training prediction entropy of the FFNN model approaches that of the RNN model, the gap between the attack accuracy of the two MIAs narrows. As a result, the ability of RNNs to maintain a lower prediction entropy than FFNNs vis-\'a-vis member data records may render them more vulnerable to MIAs.

In the next experiment, we demonstrate that the MIAs are indeed sensitive to the entropy of the victim's predictions. To illustrate this, for each inference made by the MIA, we measure the victim's cross-entropy loss—as a performance measure—and we measure the prediction entropy of the victim. Then, we generate a scatter plot in which the y-axis represents cross-entropy and the x-axis measures prediction entropy. We use two colors to distinguish between member and non-member inferences made by the MIA and use a distinct marker to represent erroneous inferences. The results in Figure \ref{fig:entropy_scatter} suggest that the MIAs divide scatter area into 4 quadrant and label data records with cross-entropy loss below a certain threshold and entropy below a certain threshold as member data. 

\paragraph{\textbf{Model Memorization as a Vulnerability Factor:}}
In the last experiment, we demonstrate that the RNN and FFNN models also differ in the way they retain their performance with respect to member data post training. 
If a model responds to a post-training query using a member data with the same accuracy as it previously held while being trained with that member data, we say that the model has \textit{memorized} its training data. If the model's accuracy for such a query decreases after training, we say that the model has \textit{forgotten} the training data.

\begin{figure*}[!ht]
    \centering
    \input{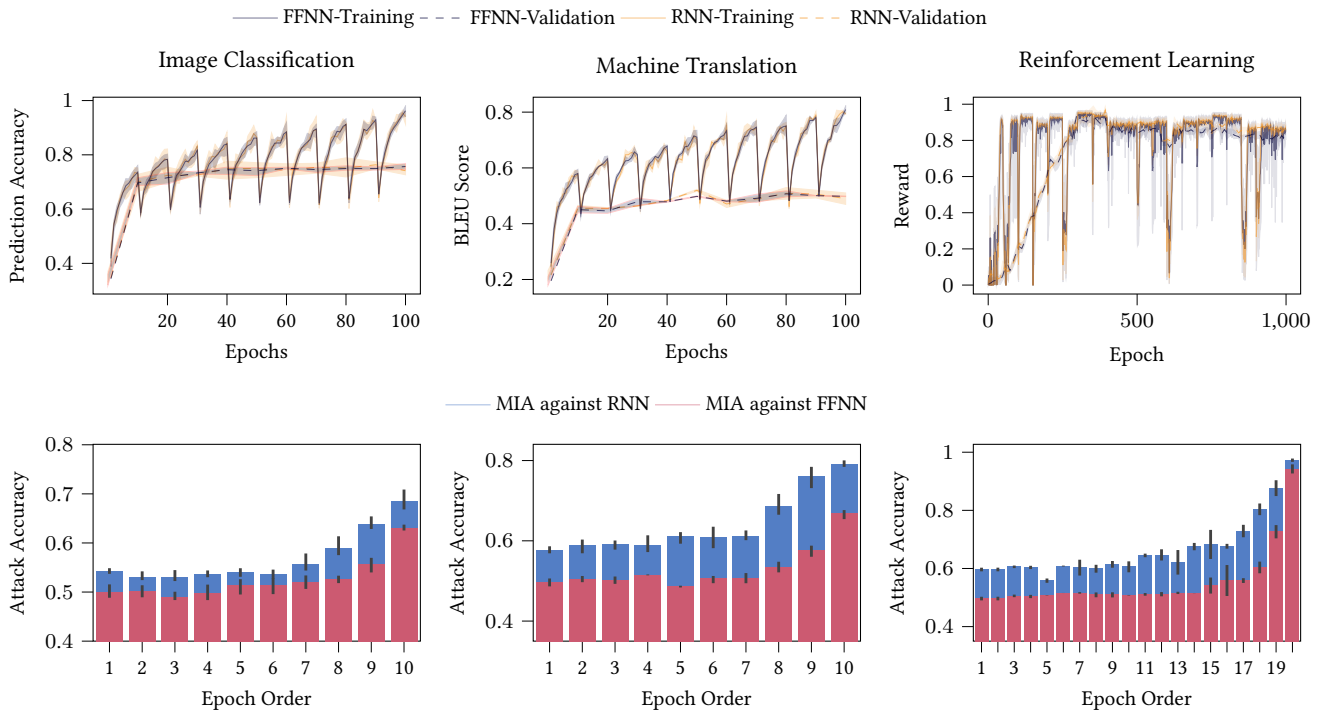}
    \caption{Comparing model memorization in RNNs and FFNNs. The top row plots the training and validation performance of the models when trained sequentially with a collection of ordered batches of training data. The saw-tooth pattern in the training performance is common in sequential training of machine learning models and is due to the \textit{catastrophic interference} phenomenon \cite{mccloskey1989catastrophic}. The validation lines are smoothed out by measuring performance at the end of every 10 epochs. The bottom row plots the attack accuracy of the MIAs with respect to the individual epochs in the order at which they were introduced to the victims during training.}
    \label{fig:sequential}
\end{figure*}

Model memorization, if not associated with overfitting, is favorable from a performance-maximizing perspective. 
For example, in the deep reinforcement learning experiment, both the RNN and the FFNN agents reach the goal state within roughly $35$ time-steps when validated in unseen floor-maps. However, when the RNN agent is queried in a member floor-map, it reaches the goal in approximately $20$ time-steps, whereas the FFNN agent still reaches the goal in $35$ time-steps. As a result, the RNN agent appears to memorize the floor-maps whereas the FFNN agent seems to forget.
We note that the reward function used in the training of the agents is relatively insensitive to the number of steps taken to reach the goal. Instead, it is more sensitive to whether or not the agent reaches the goal at all in an episode.
In particular, a $75\%$ increase in the number of steps from $20$ to $35$ decreases the total reward only by $7.42\%$ according to \eqref{eq: RL reward function}.
Hence, the RNN agent appears to memorize the floor-maps even though it was not specifically incentivized by the reward system to do so.

From a privacy perspective, such discrepancies between a model performance's with respect to seen and unseen data are harmful as they can be exploited by an adversary via MIAs. To illustrate this, we partition the training datasets into a collection of disjoint batches of data and assign an order to each batch arbitrarily at random. We then use these batches sequentially to train the RNN and FFNN models. Once the two models are trained, we report the accuracy of the MIAs with respect to percentage of correct inferences vis-\'a-vis each batch. The results in Figure \ref{fig:sequential} indicate that the MIAs' accuracy for older batches of data in FFNNs quickly diminishes to $50\%$, whereas in RNNs, the MIAs maintain non-trivial accuracy even for the early batches of data.
As a result, we posit that model memorization is another factor that renders RNNs more vulnerable to MIAs than FFNNs.

\begin{figure*}[!ht]
    \centering
    \input{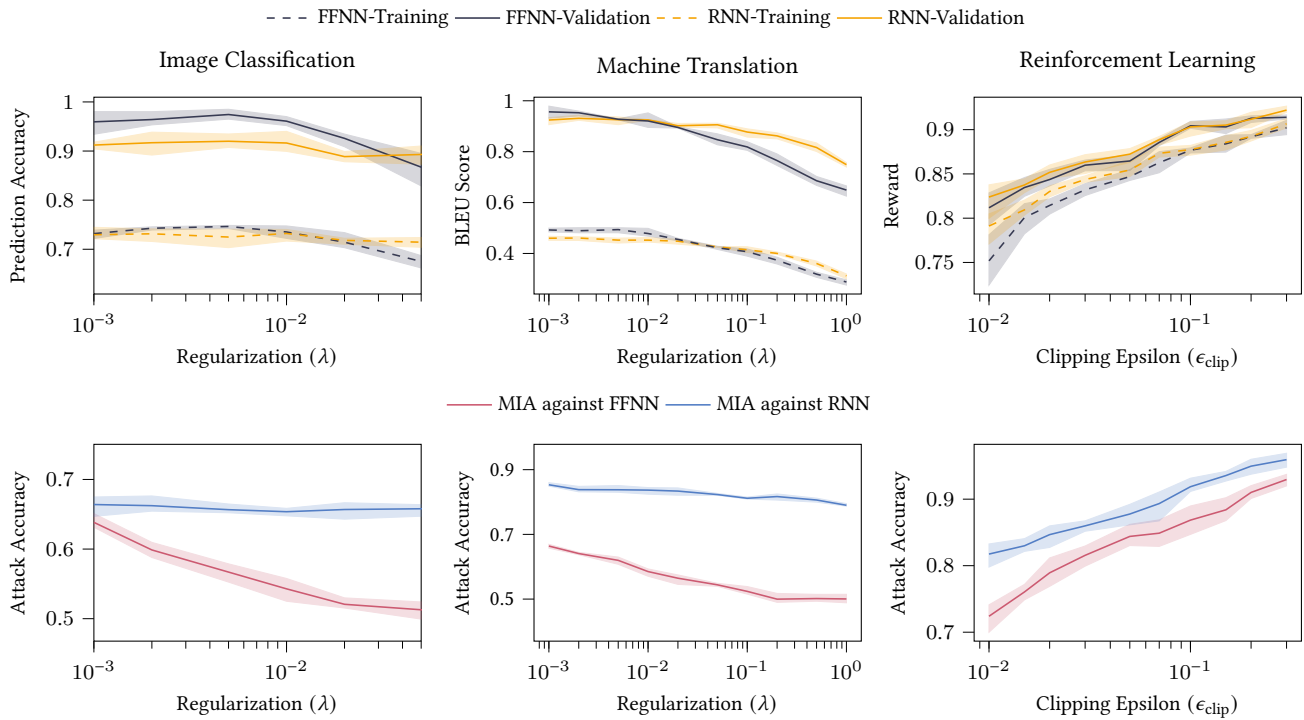}
    \caption{Effects of regularization on attack accuracy. From the left to the right column: image classification, machine translation, and deep reinforcement learning. The MIAs are trained separately for each regularization value tested and the shadow models use the same parameters as their victims.}
    \label{fig: regularization}
\end{figure*}

\section{Preempting Privacy Threats} \label{sec: DP}
In this section, we shift the focus from studying vulnerability to studying defense methods against MIAs. We first briefly discuss regularization methods, then, we study methods that leverage the promise of differential privacy.

\subsection{Defense via Regularization}
We now investigate the effects of overtraining and regularization in the considered machine learning tasks. Increasing the training time of machine learning algorithms often results in overfitting.
For example, the validation performance of the FFNN model in the image classification task decreases after training for 10 epochs in Figure \ref{fig:MIA accuracy}, whereas its training performance keeps increasing.
On the other hand, training machine learning models for an extended number of epochs may not always lead to overfitting. Such a phenomenon in RNNs was first reported in \cite{song2019auditing} for natural language processing models which is consistent with our results in Figures \ref{fig:MIA accuracy}.

Regularization methods such as $\ell_2$-regularization are effective in preventing overfitting and they have been shown to be effective in reducing the vulnerability of FFNN image-classification models to MIAs \cite{Shokri_2017, salem2018ml}. However, regularization may add bias to the converging performance levels because they alter the objective function. In particular, these methods compute the $\ell_2$-norm of the node activations as a penalty term, which is subsequently multiplied with a regularization coefficient $\lambda$ and added to the model's loss function.
In Figure \ref{fig: regularization}, we observe that regularization affects the FFNN and RNN models in the image classification and machine translation experiments differently. In particular, the MIA accuracy in the FFNN models is highly sensitive to the regularization coefficient $\lambda$, whereas the MIA accuracy against RNN models are impacted by regularization only marginally.

For the deep reinforcement learning agents, we test a different method of regularization.
It is common in deep reinforcement learning algorithms such as the PPO and trust-region policy optimization (TRPO) \cite{schulman2015trust} to regularize the Kullback-Leibler divergence between the policy updates in order to increase model stability \cite{liu2019neural}.
In the PPO algorithm, which we use to train the RNN and FFNN agents in the deep reinforcement learning experiment, a parameter called the clipping epsilon $\epsilon_\text{clip}$ controls the policy updates as follows: a small value of $\epsilon_\text{clip}$ prevents the agent from taking large gradient steps whereas a large epsilon does not restrict the agent as much. In this case, the validation performance of both the RNN and FFNN agents are sensitive to regularization. However, the RNN agent remains more vulnerable to the MIA than the FFNN agent, and its respective MIA accuracy is relatively less sensitive to regularization based on the corresponding line slopes.

\subsection{Defense via Differential Privacy}

Differential privacy is a characteristic of an algorithm and provides a quantitative definition to data privacy \cite{diff_prav2}.
A differentially private algorithm makes it hard for any observer to link the algorithm's outputs to the individual entries of the dataset that contributed to generating that output. 
It is best justified to use differential privacy when the purpose of the algorithm is to compute some aggregate information about a dataset whose entries contain sensitive information. For example, the US Census Bureau uses differential privacy to protect the data subjects in its publications \citep{abowd2018us}.
Differential privacy is formally defined as follows:
\begin{definition} \label{def:differential_privacy}
Let $f:\mathcal{D} \mapsto \mathcal{R}$ be a query function from an input domain $\mathcal{D}$ to an output domain $\mathcal{R}$.
Define two datasets $D$ and $D'$—both in $\mathcal{D}$—adjacent if the number of the entries in which the two datasets hold different values is at most one. 
Let $(\Omega,\mathcal{F},\mu)$ be a probability space and $\mathcal{M}$ be a $\sigma$-algebra such that $(\mathcal{R},\mathcal{M})$ is measurable. For a given $\epsilon\ge0$ and $\delta\in[01]$, a mechanism $M:\mathcal{R}\times \Omega \mapsto \mathcal{R}$ satisfies $(\epsilon,\delta)$-differenetial privacy if, for all $R\subseteq \mathcal{R}$ and all adjacent $D$ and $D'$,
\begin{equation} \label{eq:diff privacy}
    \mathrm{P}\left[M\left(f(D)\right)\in R\right] \le
    \exp(\epsilon) \cdot \mathrm{P}\left[M\left(f\left(D'\right)\right)\in R\right] + \delta.
\end{equation} 
\end{definition}

If a mechanism satisfies \eqref{eq:diff privacy} with $\delta=0$, it satisfies pure $\epsilon$-differential privacy. Intuitively, the parameter $\epsilon$ captures the strength of privacy protections and $\delta$ captures the probability that pure $\epsilon$-differential privacy fails. Privacy failure could happen due to two reasons: either \eqref{eq:diff privacy} holds for a larger $\epsilon$ or no finite $\epsilon$ ever satisfies pure differential privacy. It is customary to choose single-digit values for $\epsilon$ and choose $\delta$ to be $\mathcal{O}\left(|D|^{-1}\right)$, where $|D|$ is the size of the dataset that we wish to protect \cite{diff_prav2}. However, in some applications, even large values for $\epsilon$ may still provide a strong privacy shield \cite{bhowmick2018protection}.

Differential privacy is immune to post-processing, meaning that post-hoc computations on the output of a differentially private mechanism does not affect the level of differential privacy. Subsequent queries from the output of a differentially private mechanism may weaken privacy, however. In general, the overall privacy level of a sequence of $k$ queries from an $(\epsilon,\delta)$-differentially private mechanism results in $(k\epsilon, k\delta)$-differential privacy according to the Composition Theorem \cite{diff_prav2}. The overall privacy level is often referred to as the \textit{privacy budget}.
In applications wherein multiple queries are made from some sensitive dataset, one must be mindful of the total privacy budget expended.

\subsubsection{Enforcing Differential Privacy}
The methods that we use in this section to enforce differential privacy utilize the Gaussian mechanism for differential privacy. The mechanism adds a zero-mean Gaussian noise to the output of a query function with a sensitive input dataset. The mechanism calibrates the variance of the noise based of the \textit{sensitivity} of the query function, defined as follows:
\begin{definition}\label{def: sensitivity}
Let $f:\mathcal{D}\mapsto \mathcal{R}$ be a query function that maps from a dataset domain $\mathcal{D}$ to a normed space $\mathcal{R}$. The sensitivity of $f$, denoted $\mathrm{S}(f)$, is
\begin{equation}
    \mathrm{S}(f) := \sup_{D,D'} \left\|f(D) - f\left(D'\right)\right\|,
\end{equation}
where $\|\cdot\|$ is the norm operator and $D$ and $D'$ are any two adjacent datasets under the definition of adjacency established in Definition \ref{def:differential_privacy}.
\end{definition}

The following theorem from \cite{dong2019gaussian}—see Section 2.4 therein—establishes the $(\epsilon,\delta)$-differential privacy of the Gaussian mechanism.
\begin{theorem}\label{Thm:gaussian_mechanism}
Let $f$ be a query function with sensitivity $\mathrm{S}(f)$. Fix $\sigma>0$ and define the Gaussian mechanism as $M_\text{G}(f(D); \sigma) = f(D) + \xi$ such that $\xi\sim \mathcal{N}\left(0, \sigma^2\mathbf{I}\right)$. For all $\epsilon>0$, let
\begin{equation} \label{eq: dp_of_gaussian_mechanism}
    \delta = \Phi\left(-\frac{\epsilon}{\mu} + \frac{\mu}{2}\right) - \exp(\epsilon) \Phi\left(-\frac{\epsilon}{\mu}-\frac{\mu}{2}\right),
\end{equation}
where $\mu = \frac{\mathrm{S}(f)}{\sigma}$ and $\Phi$ is the cumulative distribution function of the standard normal distribution. Then, the Gaussian mechanism satisfies $(\epsilon,\delta)$-differential privacy.
\end{theorem}

Later in the experiments of this section, we deploy the Gaussian mechanism in two algorithms: the \textsc{DP-SGD} algorithm \cite{abadi2016deep} in which the Gaussian mechanism is used to privatize the gradients during the training of a neural network and a post-training privacy mechanism in which we deploy the Gaussian mechanism to privatize the trained parameters of a neural network.

\textsc{DP-SGD} modifies the stochastic gradient descent (\textsc{SGD}) algorithm such that the training algorithm itself satisfies differential privacy. In particular, the mechanism $M$ in Definition \ref{def:differential_privacy} is the training algorithm that maps a training dataset to a set of network parameters. The mechanism $M$ repeatedly performs the following at every update step:
clips the gradients computed over a batch of training data, averages the clipped gradients, invokes the Gaussian mechanism to privatize the gradients, and finally performs an SGD update with the privatized gradient. In other words, \textsc{DP-SGD} repeatedly applies the following update rule:
\begin{equation}
    \text{DP-SGD:} \quad \theta_{i+1} = \theta_i - \eta \frac{1}{|X_i|}M_\text{G}\left(\sum_{x\in X_i}\mathrm{CL}(g_x(\theta_i), C); C\sigma\right),
\end{equation}
where $i$ is the current iteration number and $\theta_i$ is the neural network's trainable parameters at iteration $i$; $\eta$ is the learning rate; $X_i$ is a minibatch of training data; and with $\ell$ the loss function and $C$ a fixed scalar,
\begin{equation}
    g_x(\theta_i) := \frac{\partial \ell(\theta, x)}{\partial \theta } \Big|_{\theta = \theta_i} \quad \text{and} \quad \mathrm{CL}(g_x, C) := g_x \cdot \min\left(1, \frac{C}{\|g_x\|}\right)
\end{equation}
are the calculated gradient and the clipping function, respectively.

\textsc{DP-SGD} comprises a \textit{moments accountant} subroutine that tracks the total privacy budget expended during training. 
The predictions that the resulting neural network subsequently generates post training preserve differential privacy with the same privacy budget because (i) differential privacy is immune to post-processing and (ii) the privatized gradients fully characterize the trained neural network given a fixed initialization $\theta_0$.

\textsc{DP-SGD} invokes the Gaussian mechanism at every gradient update step; therefore, subsequent gradient updates can mitigate the negative impacts of injecting noise on the model's utility. 
However, some queries may require more precision—or less privacy—than others. In order to adjust the level of privacy in \textsc{DP-SGD}, the weights must be retrained from scratch, which can be computationally expensive.
As a result, the \textsc{DP-SGD} algorithm may only be suitable for applications in which the underlying privacy interests necessitate \textit{limiting} the flow of information about the training data, as opposed to those necessitating a discretionary \textit{control} over the flow of such information. 

\SetKwComment{Comment}{/* }{*/ }

\begin{algorithm}[t]
\caption{\textsc{Gaussian Privacy Module}}\label{algo: mechanism}
\KwIn{hyperparameters $\mathcal{H}$ and training algorithm $\mathcal{A_H}$,
		training dataset $D$,
		Gaussian mechanism variance $\sigma$,
		query set $X$}
\KwOut{$\tilde Y$}
$\theta \leftarrow \mathcal{A_\mathcal{H}}(D)$ \Comment*[r]{Train $\mathrm{NN}_\theta$}
$\tilde \theta \leftarrow M_\text{G}(\theta;\sigma)$ \Comment*[r]{Invoke the Gaussian mechanism}
\For{$x_i$ in $X$}{
    $\tilde y_i \leftarrow \mathrm{NN}_{\tilde \theta} (x_i)$ \Comment*[r]{Respond to each query}
}
$\tilde Y \leftarrow \{y_i, i=1\dots|X|\}$
\end{algorithm}

A post-training privacy mechanism that mounts on a fully trained model as an external module can offer the flexibility required for controlling the flow of information. In this case, instead of having to retrain the model, one can apply changes to the privacy module. In Algorithm \ref{algo: mechanism}, we introduce the Gaussian privacy module (GPM) which is our proposed post-training privacy mechanism. By using the GPM, adjusting the level of privacy becomes as simple as a one-time adjustment of the variance of the Gaussian mechanism. 

We now reconcile Algorithm \ref{algo: mechanism} and Theorem \ref{Thm:gaussian_mechanism} to compute the privacy budget that the GPM consumes. The first step of the algorithm—where the weights are calculated by the training algorithm $\mathcal{A_H}$—characterizes the query function $f$ in Theorem \ref{Thm:gaussian_mechanism}. 
In order to compute the privacy parameters according to \eqref{eq: dp_of_gaussian_mechanism}, one must know the sensitivity of the query function, $\mathrm{S}(f)$, a priori. The training algorithm maps a training dataset to a set of network parameters and its sensitivity captures the extent to which adjacent training datasets generate different parameters. Without any restricting measures, the sensitivity can be arbitrarily large. The \textsc{DP-SGD} algorithm faces the same issue of unbounded sensitivity and uses gradient clipping to limit sensitivity. Inspired by the gradient-clipping trick to bound sensitivity in \text{DP-SGD}, by the following theorem, we establish an upper bound on the sensitivity of Algorithm \ref{algo: mechanism} when the training algorithm used is \textsc{SGD} with gradient clipping and loss-function smoothing.

\begin{theorem} \label{thm: bounded_sensitivity}
With $\mathcal{H}$ a fixed set of hyperparameters, including a fixed initialization and a fixed seed for generating random numbers, let $\mathcal{A_H}$ be an \textsc{SGD} algorithm modified with gradient clipping and loss-function smoothing; that is, at every iteration $i$,
\begin{equation}
    \theta_{i+1} = \theta_{i} - \eta \frac{1}{|X_i|} \sum_{x\in X_i} \mathrm{CL}\left(\mathrm{E}_{Z\sim\mathcal{N}\left(0, \sigma_\text{s}^2 \mathbf{I}\right)}\left[g_x(\theta_i + Z)\right], C\right),
\end{equation}
where $\sigma_\text{s}^2$ is the smoothing variance. 
Let the loss function $\ell$ be $L$-Lipschitz, $m$ be the minibatch size, and $\beta = L/\sigma_\text{s}$. Then, after training for $T$ iterations, it holds that 
\begin{equation}\label{eq: sensitivity}
    \mathrm{S}(\mathcal{A_H}) \le 2 \frac{\left(1 + \eta \beta\right)^T - 1}{(m-1)\beta} C.
\end{equation}
\begin{proof}
See Appendix \ref{sec:proof of theorem}.
\end{proof}
\end{theorem}

The bounded sensitivity established by Theorem \ref{thm: bounded_sensitivity} immediately implies the differential privacy of the GPM due to Theorem \ref{Thm:gaussian_mechanism}. However, the upper bound in \eqref{eq: sensitivity} grows exponentially with the training horizon $T$. It is often the case that upper bounds for sensitivity are too loose and empirical measurements of the sensitivity take much smaller values. The \textsc{SensitivitySampler} algorithm \cite{rubinstein2017pain} in combination with the notion of \textit{random differential privacy} \cite{Hall_Wasserman_Rinaldo_2013} address such an issue. The former is an algorithm that estimates sensitivity and the latter is a relaxation of $(\epsilon,\delta)$-differential privacy.

\begin{definition} \label{def:gamma_differential_privacy}
The mechanism $M$ in Definition \ref{def:differential_privacy} satisfies $(\epsilon,\delta)$-random differential privacy with confidence $\gamma\in(0,1)$ if, for all adjacent datasets $D$ and $D'$ drawn from a fixed data source $\mathrm{DS}$,
\begin{multline}\label{eq:diff privacy 2}
    \mathop{\mathrm{P}}\limits_{\substack{D\sim\text{DS} \\ D'\sim\text{DS}}}\left[\forall R \subset \mathcal{R}, \mathop{\mathrm{P}}\limits_{y\sim M\left(f(D)\right)}\left[y\in R\right] \le \right. \\ \left. \exp(\epsilon) \cdot \mathop{\mathrm{P}}\limits_{y'\sim M\left(f(D')\right)}\left[y'\in R\right] + \delta \right] 
         \ge 1 - \gamma.
\end{multline}
\end{definition}

Compared to $(\epsilon,\delta)$-differential privacy wherein $\delta$ captures the probability of privacy failure due to unlikely \textit{outputs}, random differential privacy considers $\gamma$ as the probability that $(\epsilon,\delta)$-differential privacy fails due to unlikely \textit{input} datasets \cite{rubinstein2017pain}.

\begin{algorithm}[t]
\caption{\textsc{SensitivitySampler}}\label{algo:Sensitivity Sampler}
\KwIn{training algorithm $\mathcal{A_H}$ with
		hyperparameters $\mathcal{H}$,
		sample size $n$,
		data source $\mathrm{DS}$,
		training dataset size $N$,
		}
\KwOut{$\bar S$}

\For{$i=1$ to $n$}{
\For{$j=1$ to $N+1$}{
$d_j\sim \mathrm{DS}$ \Comment*[r]{Sample from the data source}
}
$D_1 \leftarrow \cup_{j=1}^N d_j$ \;
$D_2 \leftarrow \left(\cup_{j=1}^{N+1} d_j\right)\setminus \{d_N\}$\;
$\theta \leftarrow \mathcal{A_H}(D_1)$ \Comment*[r]{Train Model 1}
$\theta' \leftarrow \mathcal{A_H}(D_2)$ \Comment*[r]{Train Model 2}
$\bar S^{(i)} \leftarrow \|\theta - \theta'\|_2$\;
}
$\bar S \leftarrow  \max_i \bar S^{(i)}$\;
\end{algorithm}

We use the \textsc{SensitivitySampler} algorithm in the context of training a neural network for machine learning as described in Algorithm \ref{algo:Sensitivity Sampler}. The algorithm repeatedly samples two adjacent training datasets from a fixed data source, invokes the training algorithm for both of the sampled training datasets, and estimates the sensitivity of the training algorithm based on the maximum 2-norm difference between the observed network parameters. The following theorem, which is an immediate result of Corollary 20 of \cite{rubinstein2017pain}, establishes the random differential privacy of the GPM.

\begin{theorem}
\label{thm:sample_dp} Fix a set of hyperparameters $\mathcal{H}$ and training algorithm $\mathcal{A_H}$. Let $\bar S$ be the output of Algorithm \ref{algo:Sensitivity Sampler} run with $n$ samples. Further, let 
\begin{equation} \label{eq: rho_gamma}
    \rho = \exp\left(\frac{1}{2} W_{-1}\left(-\frac{1}{4n}\right)\right) \quad \text{and} \quad
        \gamma  = \rho + \sqrt{\frac{\log(1/\rho)}{2n}},
\end{equation}
where $W_{-1}$ is the Lambert W function defined as the inverse relation of the function $f(z) = z\exp(z)$. With $\sigma$ the variance of the Gaussian mechanism in Algorithm \ref{algo: mechanism}, for all $\epsilon>0$ and
\begin{equation}
\delta = \Phi\left(-\frac{\epsilon}{\mu} + \frac{\mu}{2}\right) - \exp(\epsilon) \Phi\left(-\frac{\epsilon}{\mu}-\frac{\mu}{2}\right),
\end{equation}
where $\mu = \bar S/\sigma$, Algorithm \ref{algo: mechanism} satisfies $(\epsilon,\delta)$-random differential privacy with confidence $\gamma$.
\end{theorem}

With the theoretical preliminaries set in this subsection, we now move on to the experiments.


\subsubsection{Experiments}

Similar to Section \ref{sec: vuln} in which we compared vulnerability to MIAs, we consider RNN and FFNN models in three representative machine learning tasks, namely image classification, machine translation, and deep reinforcement learning. However, for the machine translation task, we fine-tune a pre-trained model, BERT \cite{devlin-etal-2019-bert}, with a subset of training samples from the \texttt{WMT14} English-French training dataset \cite{Bojar2014FindingsOT} instead of training a model from scratch using the \texttt{Multi30K} dataset. \texttt{WMT14} contains substantially more samples than \texttt{Multi30K} and is therefore more suitable for the \textsc{SensitivitySampler} algorithm.

In the first experiment, we use the \textsc{DP-SGD} algorithm to enforce differential privacy using a range of values for noise variance. Then, we measure the cost of privacy in terms of utility loss, which we formally define as follows:
\begin{definition}
    \label{def:util_loss}
    Let $\mathbb{M}$ be an evaluation metric that takes as input a set of predictions $Y$ alongside their ground-truth labels $Y_\text{GT}$, and returns a numerical value that indicates the quality of the predictions. Then, the utility loss is
    \begin{center}
        $\mathcal{L}_{util} = 1 - \frac{\mathbb{M}(\tilde Y, Y_\text{GT})}{\mathbb{M}(Y, Y_\text{GT})}$.
    \end{center}
\end{definition}

\begin{figure*}[t]
    \centering
    \input{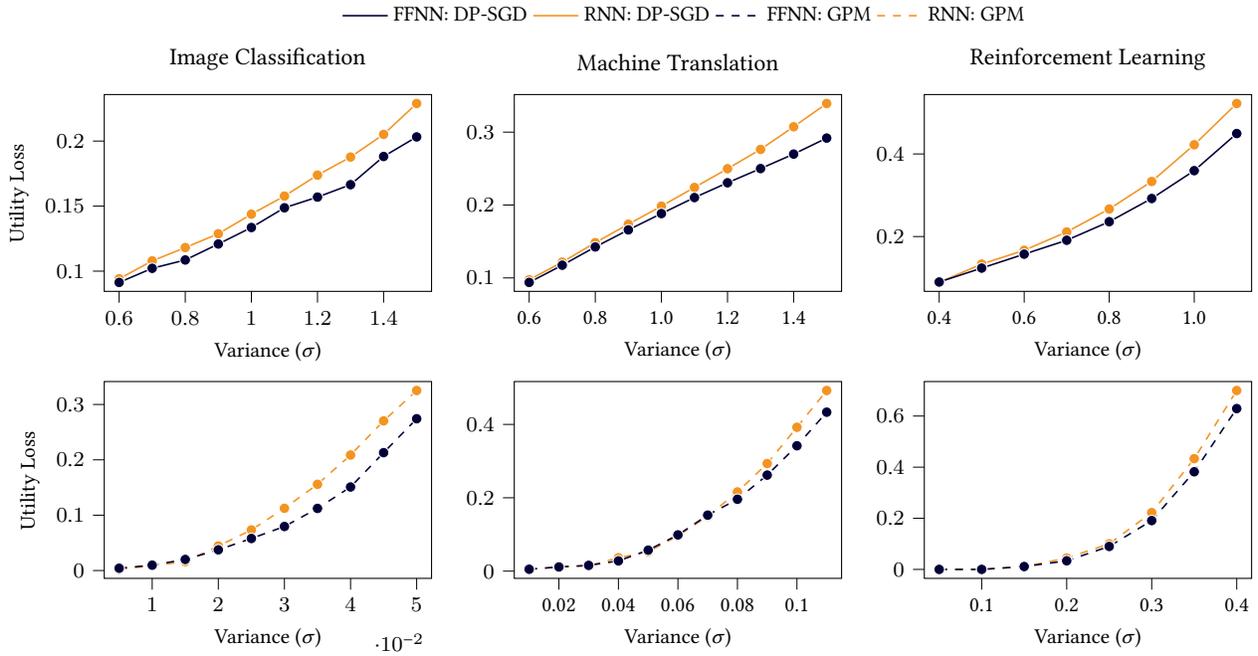}
    \caption{The privacy-utility trade-off of DP-SGD and GPM. The top row corresponds to the utility loss caused by DP-SGD and the second row corresponds to the utility loss caused by the GPM. The RNNs consistently trade off more utility than FFNNs for both DP-SGD and GPM.
    }
    \label{fig:noise_util}
\end{figure*}

We now report the results. The top row of Figure \ref{fig:noise_util} indicates that the RNN models consistently trade off more utility than the FFNN models at every noise variance tested.
The same level of noise translates to the same level of $(\epsilon,\delta)$-differential privacy in \textsc{DP-SGD};
as a result, enforcing the same level of $(\epsilon,\delta)$-differential privacy is more costly in RNNs than FFNNs with respect to utility loss.

A similar observation can be made when the GPM enforces random differential privacy for the RNN and FFNN models.
In this experiment, we fine-tune the hyperparameters of the training algorithms such that: (i) the two models achieve similar validation performance levels before the GPM is deployed and (ii) Algorithm \ref{algo:Sensitivity Sampler} estimates the same level of sensitivity for the two models as reported in Table \ref{tab:util_spec}. We refer to these estimates as \textit{empirical sensitivity}.
The empirical sensitivities in Table \ref{tab:util_spec} correspond to $n=500$ samples which translates to confidence $\gamma < 0.08$ established by \eqref{eq: rho_gamma} in Theorem \ref{thm:sample_dp}. 

\begin{table}[!h]
    \centering
    \caption{Empirical sensitivity estimated by Algorithm \ref{algo:Sensitivity Sampler}.}
    \begin{tabular}{|c||c|c||}
         \hline
         Task & $\bar{S}_\text{RNN}$ & $\bar{S}_\text{FFNN}$ \\
         \hline
         Image Classification &  0.013209 & 0.013518\\
         Machine Translation & 0.11678 & 0.11845\\
         Reinforcement Learning & 0.093682 & 0.093429\\
         \hline
    \end{tabular}
    \label{tab:util_spec}
\end{table}

In Figure \ref{fig:noise_util}, where we plot utility loss vs. noise variance, it can be observed that deploying the GPM consistently trades off more utility in RNNs than FFNNs.
The results in Figure \ref{fig:noise_util} also illustrate that the RNNs trade off more utility for the same level of random differential privacy because the sensitivities of the two models are approximately equal.

\section{Conclusion} \label{sec: conc}

In this work, we provided empirical evidence that MIAs can achieve higher accuracy when they attack RNNs compared with their FFNN counterparts.
We showed that RNNs maintain a larger entropy gap between the predictions corresponding to member data and those corresponding to unseen data as a key vulnerability factor that is more elevated in RNNs than FFNNs.
We also found that RNNs memorize their training data in a way that an MIA can maintain a non-trivial attack accuracy over the entire history of their training, whereas the corresponding attack accuracy for the FFNNs quickly drops to 50\% as we move back in the training history. 

In the second part of the study, we considered two prominent mitigation methods: weight regularization and differential privacy. Then, we showed that regularization was less effective in protecting RNNs compared to FFNNs. 
Moreover, we showed that enforcing differential privacy in RNNs can be more costly than FFNNs in terms of the privacy-utility trade-off.

We conclude this paper with the observation that the privacy risks of deploying RNNs in machine learning are higher than FFNNs with the same level of performance.
Alongside the existing computational drawbacks of training RNNs, our results provide further incentives to replace RNNs with FFNNs.


\bibliographystyle{ACM-Reference-Format}
\bibliography{refs}

\newpage
\appendix
\section{Appendix}
\subsection{Proof of Theorem \ref{thm: bounded_sensitivity}} \label{sec:proof of theorem}
\begin{theorem} 
With $\mathcal{H}$ a fixed set of hyperparameters, including a fixed initialization and a fixed seed for generating random numbers, let $\mathcal{A_H}$ be an \textsc{SGD} algorithm modified with gradient clipping and loss-function smoothing; that is, at every iteration $i$,
\begin{equation}
    \theta_{i+1} = \theta_{i} - \eta \frac{1}{|X_i|} \sum_{x\in X_i} \mathrm{CL}\left(\mathrm{E}_{Z\sim\mathcal{N}\left(0, \sigma_\text{s}^2 \mathbf{I}\right)}\left[g_x(\theta_i + Z)\right], C\right),
\end{equation}
where $\sigma_\text{s}^2$ is the smoothing variance. 
Let the loss function $\ell$ be $L$-Lipschitz, $m$ be the minibatch size, and $\beta = L/\sigma_\text{s}$. Then, after training for $T$ iterations, it holds that 
\begin{equation}\label{eq: sensitivity_}
    \mathrm{S}(\mathcal{A_H}) \le 2 \frac{\left(1 + \eta \beta\right)^T - 1}{(m-1)\beta} C.
\end{equation}
\begin{proof}
Let $\bar\ell(\theta_i, x) = \mathrm{E}_{Z\sim\mathcal{N}\left(0, \sigma_\text{s}^2 \mathbf{I}\right)}\left[\ell(\theta_i + Z, x)\right]$. Such an operation is known as randomized smoothing which transforms the $L$-Lipschitz loss function $\ell$ into $L/\sigma_\text{s}$-smooth $\bar \ell$ \cite{scaman2020simple}; that is,
\begin{equation}
    \left\|\frac{\partial \bar\ell(\theta, x)}{\partial \theta} \Big|_{\theta = a} - \frac{\partial \bar\ell(\theta, x)}{\partial \theta} \Big|_{\theta = b} \right\| \le \frac{L}{\sigma_\text{s}} \|a - b\|.
\end{equation}
We also have that
\begin{equation}
    \frac{\partial \bar\ell(\theta, x)}{\partial \theta } \Big|_{\theta = \theta_i} = \mathrm{E}_{Z\sim\mathcal{N}\left(0, \sigma_\text{s}^2 \mathbf{I}\right)}\left[g_x(\theta_i + Z)\right].
\end{equation}
Considering SGD's update rule with clipped gradients and randomized smoothing, we have that, for two adjacent datasets $D$ and $D'$ and their respective minibatches at stage 0, $X_0$ and $X'_0$,
\begin{equation} \label{eq: stage1}
    \theta_1 = \theta_0 - \eta \frac{1}{|X_0|} \sum_{x\in X_0} \mathrm{CL}\left(\frac{\partial \bar\ell(\theta, x)}{\partial \theta} \Big|_{\theta_0}, C\right) 
\end{equation}
and
\begin{equation}
    \theta'_1 = \theta_0 - \eta \frac{1}{|X'_0|} \sum_{x'\in X'_0} \mathrm{CL}\left(\frac{\partial \bar\ell(\theta, x')}{\partial \theta} \Big|_{\theta_0}, C\right).
\end{equation}
The two minibatches can only differ in one data record and fixing the random seeds ensures that the same data indices will be chosen for both $X_0$ and $X'_0$. As a result,
\begin{multline} 
    \|\theta_1 - \theta'_1\| = \\
    \eta \left\|\sum_{x\in X_0 \setminus X'_0} \mathrm{CL}\left(\frac{\partial \bar\ell(\theta, x)}{\partial \theta} \Big|_{\theta_0}, C\right)- \sum_{x'\in X'_0 \setminus X_0} \mathrm{CL}\left(\frac{\partial \bar\ell(\theta, x')}{\partial \theta} \Big|_{\theta_0}, C\right) \right\| \\
    \le 2 \eta \frac{ C}{m}.
\end{multline}
For the next SGD update, we write
\begin{equation}
        \theta_2 = \theta_1 - \eta \frac{1}{|X_1|} \sum_{x\in X_1} \mathrm{CL}\left(\frac{\partial \bar\ell(\theta, x)}{\partial \theta} \Big|_{\theta_1}, C\right)
\end{equation}
and
\begin{equation}
     \theta'_2 = \theta'_1 - \eta \frac{1}{|X'_1|} \sum_{x'\in X'_1} \mathrm{CL}\left(\frac{\partial \bar\ell(\theta, x')}{\partial \theta} \Big|_{\theta'_1}, C\right).
\end{equation}
Due to the smoothness of $\bar\ell$, we have that
\begin{multline}
    \left\|\mathrm{CL}\left(\frac{\partial \bar\ell(\theta, x)}{\partial \theta} \Big|_{\theta = a},C\right) - \mathrm{CL}\left(\frac{\partial \bar\ell(\theta, x)}{\partial \theta} \Big|_{\theta = b}, C\right) \right\| \le \\ \min\left(2C,\frac{L}{\sigma_\text{s}} \|a - b\|\right).
\end{multline}
With $\beta = \frac{L}{\sigma_\text{s}}$, we can write
\begin{multline}\label{eq:24}
    \|\theta_2 - \theta'_2\| \le \|\theta_1 - \theta'_1\| + \eta \left(1-\frac{1}{m}\right)\|\theta_1 - \theta'_1\| \beta + 2\eta\frac{C}{m}.
\end{multline}
The reason that \eqref{eq:24} holds is that $X_1$ and $X'_1$ are obtained from two adjacent datasets and because of the fixed-seed assumption, they hold equal entries except for one; for the equal entries, the second term on the right-hand side of \eqref{eq:24} can be used and for the non-equal entry, the third term can be used as an upper bound. Analogously, for every stage $i\ge 2$, we have
\begin{equation}
    \|\theta_i - \theta'_i\| \le 2\eta \frac{C}{m} + \left[1 + \eta \left(1-\frac{1}{m}\right)\beta\right]\|\theta_{i-1} - \theta'_{i-1}\|,
\end{equation}
or
\begin{equation}
    \|\theta_i - \theta'_i\| \le 2 \frac{\left(1 + \eta \beta\right)^i - 1}{(m-1)\beta} C,  
\end{equation}
which concludes the proof.
\end{proof}
\end{theorem}
\subsection{Reproducibility Information}
\label{sec:repro}
In this section, we state the hyperparameters that we used in the experiments.
\paragraph{MIA on the reinforcement learning agent:} 
We use the PPO algorithm to train the agents, for which we use the default parameters set by the RL-Starter-Files toolbox unless stated below. The feed-forward agent uses an MLP with two hidden layers, each of which consists of 74 neurons. The RNN agent uses the MLP architecture that consists of two 32-neurons layers with 4 additional LSTM units. The first layer is activated by $\tanh$ functions and the last layer is activated by a $\mathrm{softmax}$ function. We train the agents for a total of 204,800 iterations on seeds 1 to 16 for both agents. We use the default clipping epsilon 0.2 while training.

For the implementation of the MIA, we use an MLP with 5 ReLU-activated hidden layers and 1 LSTM unit. We use 6400 \texttt{in} trajectories and 6400 \texttt{out} trajectories to generate the binary classifier's training dataset. We train the binary classifier using the Adam optimizer and the cross-entropy loss function for 15 epochs, each of which consists of 100 gradient updates. We use the Keras library \cite{chollet2015keras} to train the binary classifier with a learning rate of 0.001 and default parameters unless stated above.


\paragraph{MIA on the machine translation model:} 
We use an LSTM encoder-decoder network with dot product attention mechanism \citep{luong2015effective} to construct the sequence-to-sequence model. We use the Multi30K dataset \citep{multi30k} which consists of 30,000 sentence pairs for training and 1,000 pairs for testing. We use 5,000 sentence pairs to train the shadow model and a negative likelihood loss to update gradients. The shadow model is trained for 20 epochs, with a word-embedding dimension of 150, a hidden dimension of 200, a learning rate of 0.001, and a dropout rate of 0.2. We use PyTorch \citep{paszke2019pytorch} to implement and train the victim model with default parameters unless specified above. Once the shadow model is fine-tuned, we use 2,000 output sequences to populate the training dataset of the MIA's binary classifier. In the training procedure, we set the max norm of the gradients to 10 and clip the gradients with norms above the threshold.

We use a transformer as the FFNN structure. The transformer architecture is identical to the model from 'attention is all your need', trained with default parameters.

The binary classifier consists of 1 LSTM unit, two linear layers, a ReLU-activated layer, and a softmax layer. We implement the MIA classifier using PyTorch and train it using the cross-entropy loss function for 20 epochs with the default parameters.

\paragraph{MIA on the image classification model:} 

We use ResNet101 \cite{He2016DeepRL} implemented in the Keras library \cite{chollet2015keras} as the FFNN model for image classification. ResNet101 consists of 101 convolutional layers followed by one max-pooling layer, one fully connected linear layer, and an output layer with softmax activation. 

We use ReNet \cite{visin2015renet} implemented by PyTorch \cite{paszke2019pytorch} under default parameters as the RNN model for image classification. ReNet consists of 4 bi-directional LSTMs, 2 fully connected layers with ReLU activation, and an output layer with softmax activation. We train both models using the categorical cross-entropy loss function as their learning objective function and use the Adam optimizer. The learning rates used are $0.001$ and $0.01$ for ResNet101 and ReNet, respectively.

We use the image classification dataset Cifar10 which consists of 50,000 training records and 10,000 testing records. We train the target model and shadow model using 10,000 training records and a categorical cross-entropy loss is used to update the gradient.  We clip the gradients whose norm is greater than 10.

For the implementation of the MIA, we use an MLP with 5 ReLU-activated hidden layers. We train the classifier using 20,000 probability pairs with half labeled `in'. We use the Keras library \cite{chollet2015keras} to train the binary classifier with a learning rate of 0.001 and default parameters unless stated above.

\begin{table*}
    \centering
    \caption{Experiment Specifications.}
    \begin{tabular}{||c|c|c|c|c|c|c|c||}
         \hline
         Task & train size $|D|$ & $\bar S_{RNN}$ & $\bar S_{FFNN}$ & $\delta$ & clip norm & batch & epoch \\
         \hline
         IC & 10000 &  0.013209 & 0.013518 & 1e-4 & 10 & 128 & 50\\
         RL & 6400(32)  & 0.093682 & 0.093429 & 1e-4 & 10 & 128 & 100 \\
         NMT & 5000 & 0.11678 & 0.11845 & 1e-4 & 10 & 128 & 50\\
         \hline
    \end{tabular}
    \label{tab:util_spec_}
\end{table*}

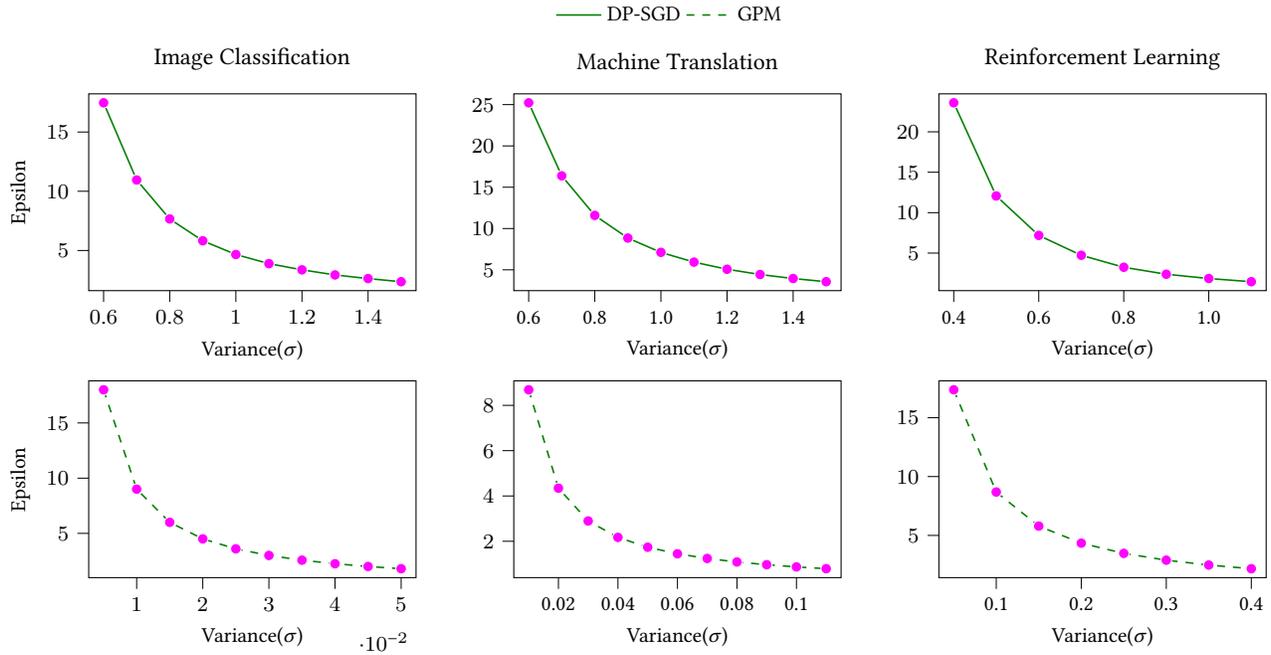
\begin{figure*}[!ht]
    \centering
    \begin{tikzpicture}

\definecolor{color0}{RGB}{255,0,255}
\definecolor{color1}{RGB}{0,128,0}

\begin{groupplot}[group style={group name = plots, group size=3 by 2, vertical sep=1.2cm, horizontal sep=1.3cm}, width=0.7\columnwidth, height=4.2cm]


\nextgroupplot[
legend cell align={left},
legend style={
  fill opacity=0,
  draw opacity=1,
  text opacity=1,
  at={(1.4, 1.3)},
  anchor=south west,
  draw=white,
  font= \small
},
legend columns = 2,
tick align=outside,
tick pos=left,
x grid style={white!69.0196078431373!black},
xlabel={Variance($\sigma$)},
xmin=0.555, xmax=1.545,
tick pos=left,
xtick = {0.6, 0.8, 1.0,1.2,1.4},
xtick style={color=black},
y grid style={white!69.0196078431373!black},
ylabel={Epsilon},
title={Image Classification},
ymin=1.61089040200206, ymax=18.2223283913991,
ytick style={color=black},
xlabel style ={font =\small},
ylabel style ={font =\small},
ticklabel style={font=\small}
]
\addplot[semithick, color1]
table {%
0 0
};
\addlegendentry{DP-SGD}

\addplot[semithick, color1, dashed]
table {%
0 0
};
\addlegendentry{GPM}

\addplot [draw=white, fill=color0, mark=*, only marks]
table{%
x  y
0.6 17.4672630282447
0.7 10.952229519813
0.8 7.6612717829645
0.9 5.82041656492383
1 4.66251178101704
1.1 3.89115409372612
1.2 3.36818211064323
1.3 2.93726769262041
1.4 2.630101914167
1.5 2.36595576515647
};
\addplot [semithick, color1]
table {%
0.6 17.4672630282447
0.7 10.952229519813
0.8 7.6612717829645
0.9 5.82041656492383
1 4.66251178101704
1.1 3.89115409372612
1.2 3.36818211064323
1.3 2.93726769262041
1.4 2.630101914167
1.5 2.36595576515647
};

\nextgroupplot[
legend cell align={left},
legend style={
  fill opacity=0.8,
  draw opacity=1,
  text opacity=1,
  at={(0.03,0.97)},
  anchor=north west,
  draw=white!80!black
},
tick align=outside,
tick pos=left,
x grid style={white!69.0196078431373!black},
xlabel={Variance($\sigma$)},
xmin=0.555, xmax=1.545,
tick pos=left,
xtick = {0.6, 0.8, 1.0,1.2,1.4},
xticklabels = {0.6, 0.8, 1.0,1.2,1.4},
xtick style={color=black},
y grid style={white!69.0196078431373!black},
title={Machine Translation},
ymin=2.47955797136645, ymax=26.3034548724274,
ytick = {5, 10, 15, 20, 25},
ytick style={color=black},
xlabel style ={font =\small},
ylabel style ={font =\small},
ticklabel style={font=\small}
]
\addplot [draw=white, fill=color0, mark=*, only marks]
table{%
x  y
0.6 25.2205504678337
0.7 16.3811409338261
0.8 11.5886094476738
0.9 8.84670906946103
1 7.11302156666922
1.1 5.93266997572847
1.2 5.06976123992031
1.3 4.43442398422172
1.4 3.94671869255769
1.5 3.56246237596013
};
\addplot [semithick, color1]
table {%
0.6 25.2205504678337
0.7 16.3811409338261
0.8 11.5886094476738
0.9 8.84670906946103
1 7.11302156666922
1.1 5.93266997572847
1.2 5.06976123992031
1.3 4.43442398422172
1.4 3.94671869255769
1.5 3.56246237596013
};

\nextgroupplot[
legend cell align={left},
legend style={
  fill opacity=0,
  draw opacity=1,
  text opacity=1,
  at={(0.1,1.1)},
  anchor=south west,
  draw=white,
  font= \small
},
legend columns = 3,
tick align=outside,
tick pos=left,
x grid style={white!69.0196078431373!black},
xlabel={Variance($\sigma$)},
xmin=0.365, xmax=1.135,
tick pos=left,
xtick = {0.4,0.6,0.8,1.0,1.2},
xticklabels = {0.4,0.6,0.8,1.0,1.2},
xtick style={color=black},
y grid style={white!69.0196078431373!black},
ymin=0.354093845357828, ymax=24.6967001511726,
ytick style={color=black},
ytick = {5, 10, 15, 20, 25},
title={Reinforcement Learning},
xlabel style ={font =\small},
ylabel style ={font =\small},
ticklabel style={font=\small}
]
\addplot [draw=white, fill=color0, mark=*, only marks]
table{%
x  y
0.4 23.5902180463629
0.5 12.0665803997555
0.6 7.18586056535871
0.7 4.73098071477022
0.8 3.24399617875876
0.9 2.38425397624529
1 1.85493346629051
1.1 1.46057595016759
};
\addplot [semithick, color1]
table {%
0.4 23.5902180463629
0.5 12.0665803997555
0.6 7.18586056535871
0.7 4.73098071477022
0.8 3.24399617875876
0.9 2.38425397624529
1 1.85493346629051
1.1 1.46057595016759
};


\nextgroupplot[
legend cell align={left},
legend style={
  fill opacity=0.8,
  draw opacity=1,
  text opacity=1,
  at={(0.03,0.97)},
  anchor=north west,
  draw=white!80!black
},
tick align=outside,
tick pos=left,
x grid style={white!69.0196078431373!black},
xlabel={Variance($\sigma$)},
xmin=0.00275, xmax=0.05225,
tick pos=left,
xtick = {0.01, 0.02,0.03,0.04,0.05},
xtick style={color=black},
y grid style={white!69.0196078431373!black},
ylabel={Epsilon},
ymin=0.990079739643587, ymax=18.8115150532282,
ytick style={color=black},
xlabel style ={font =\small},
ylabel style ={font =\small},
ticklabel style={font=\small}
]

\addplot [draw=white, fill=color0, mark=*, only marks]
table{%
x  y
0.005 18.0014498117016
0.01 9.00072490585079
0.015 6.00048327056719
0.02 4.50036245292539
0.025 3.60028996234032
0.03 3.0002416352836
0.035 2.57163568738594
0.04 2.2501812264627
0.045 2.00016109018906
0.05 1.80014498117016
};
\addplot [semithick, color1, dashed]
table {%
0.005 18.0014498117016
0.01 9.00072490585079
0.015 6.00048327056719
0.02 4.50036245292539
0.025 3.60028996234032
0.03 3.0002416352836
0.035 2.57163568738594
0.04 2.2501812264627
0.045 2.00016109018906
0.05 1.80014498117016
};

\nextgroupplot[
legend cell align={left},
legend style={
  font=\small,
  fill opacity=0.8,
  draw opacity=1,
  text opacity=1,
  at={(0.03,0.97)},
  anchor=north west,
  draw=white!80!black
},
tick align=outside,
tick pos=left,
x grid style={white!69.0196078431373!black},
xlabel={Variance($\sigma$)},
xmin=0.005, xmax=0.115,
tick pos=left,
xtick = {0.02,0.04,0.06,0.08,0.1},
xticklabels = {0.02,0.04,0.06,0.08,0.1},
xtick style={color=black},
y grid style={white!69.0196078431373!black},
ymin=0.394873845808979, ymax=9.08209845360652,
ytick = {2, 4, 6, 8},
ytick style={color=black},
xlabel style ={font =\small},
ylabel style ={font =\small},
ticklabel style={font=\small}
]
\addplot [draw=white, fill=color0, mark=*, only marks]
table{%
x  y
0.01 8.68722460779754
0.02 4.34361230389877
0.03 2.89574153593251
0.04 2.17180615194939
0.05 1.73744492155951
0.06 1.44787076796626
0.07 1.24103208682822
0.08 1.08590307597469
0.09 0.965247178644171
0.1 0.868722460779754
0.11 0.789747691617958
};

\addplot [semithick, color1, dashed]
table {%
0.01 8.68722460779754
0.02 4.34361230389877
0.03 2.89574153593251
0.04 2.17180615194939
0.05 1.73744492155951
0.06 1.44787076796626
0.07 1.24103208682822
0.08 1.08590307597469
0.09 0.965247178644171
0.1 0.868722460779754
0.11 0.789747691617958
};

\nextgroupplot[
legend cell align={left},
legend style={
  fill opacity=0.8,
  draw opacity=1,
  text opacity=1,
  at={(0.03,0.97)},
  anchor=north west,
  draw=white!80!black
},
tick align=outside,
tick pos=left,
x grid style={white!69.0196078431373!black},
xlabel={Variance($\sigma$)},
xmin=0.0325, xmax=0.4175,
tick pos=left,
xtick = {0.1,0.2,0.3,0.4,0.5},
xticklabels = {0.1,0.2,0.3,0.4,0.5},
xtick style={color=black},
y grid style={white!69.0196078431373!black},
ymin=1.4116739987671, ymax=18.1345813687774,
ytick style={color=black},
ytick = {5, 10, 15, 20},
xlabel style ={font =\small},
ylabel style ={font =\small},
ticklabel style={font=\small}
]
\addplot [draw=white, fill=color0, mark=*, only marks]
table{%
x  y
0.05 17.3744492155951
0.1 8.68722460779754
0.15 5.79148307186503
0.2 4.34361230389877
0.25 3.47488984311902
0.3 2.89574153593251
0.35 2.48206417365644
0.4 2.17180615194939
};

\addplot [semithick, color1, dashed]
table {%
0.05 17.3744492155951
0.1 8.68722460779754
0.15 5.79148307186503
0.2 4.34361230389877
0.25 3.47488984311902
0.3 2.89574153593251
0.35 2.48206417365644
0.4 2.17180615194939
};

\end{groupplot}

\end{tikzpicture}
    \caption{privacy budget $\epsilon$ at each Gaussian noise level. The first row shows the privacy level of DP-SGD and the second row shows the privacy level of proposed GPM.
    }
    \label{fig:noise_eps}
\end{figure*}

\subsection{Privacy Level vs. Noise Variance}
Figure \ref{fig:noise_eps} shows the private budget $\epsilon$ at each noise level $\sigma$. Together with Figure \ref{fig:noise_util}, we observe that the proposed GPM can achieve a high privacy level ($\epsilon < 5$) with a utility loss less than 10\%.
DP-SGD can also achieve a reasonable privacy level ($\epsilon < 10$) with a utility loss lower than 15\%.

To obtain the results, we run the DP experiments following the specifications stated in Table \ref{tab:util_spec_}.

\end{document}